%% file: main.tex
\documentclass[acmsmall]{acmart}

\pdfoutput=1

\usepackage{amsmath,amsfonts}
\usepackage{algorithmic}
\usepackage{graphicx}
\usepackage{enumitem}
\usepackage{textcomp}
\usepackage{listings}

\usepackage{longtable}
\usepackage{float}
\usepackage{graphicx}
\usepackage{multirow}
\usepackage{balance}
\usepackage{hyperref}
\usepackage{tabularx} 
\usepackage{booktabs} 
\usepackage{colortbl}
\usepackage{hhline}
\usepackage{array}
\newcolumntype{L}[1]{>{\raggedright\let\newline\\\arraybackslash\hspace{0pt}}m{#1}}
\newcolumntype{C}[1]{>{\centering\let\newline\\\arraybackslash\hspace{0pt}}m{#1}}
\newcolumntype{R}[1]{>{\raggedleft\let\newline\\\arraybackslash\hspace{0pt}}m{#1}}
\usepackage{subfigure}
\usepackage{enumitem}

\newenvironment{boxedtext}
    {
    
    \begin{center}

    \begin{tabular}{|p{0.96\linewidth}|}
    \hline
    }
    { 
    \\ \hline
    \end{tabular} 
    
    \end{center}
       }

\definecolor{dartmouthgreen}{rgb}{0.05, 0.5, 0.06}

\AtBeginDocument{%
  \providecommand\BibTeX{{%
    \normalfont B\kern-0.5em{\scshape i\kern-0.25em b}\kern-0.8em\TeX}}}

\acmJournal{TOSEM}
\acmYear{2025} \acmVolume{34} \acmNumber{0} \acmArticle{0} \acmMonth{0} 

\begin{document}
\title{Automated Identification of Sexual Orientation and Gender Identity Discriminatory Texts from Issue Comments}

\author{Sayma Sultana}
\email{sayma@wayne.edu}
\affiliation{%
  \institution{Wayne State University}
  \city{Detroit}
  \state{Michigan}
  \country{USA}
}

\author{Jaydeb Sarker}

\email{jsarker@unomaha.edu}
\affiliation{%
  \institution{University of Nebraska Omaha}
  \city{Omaha}
  \state{Nebraska}
  \country{USA}
}

\author{Farzana Israt}

\email{farzanaisrat@wayne.edu}
\affiliation{%
  \institution{Wayne State University}
  \city{Detroit}
  \state{Michigan}
  \country{USA}
}

\author{Rajshakhar Paul}

\email{rpaul@wayne.edu}
\affiliation{%
  \institution{Idaho State University}
  \city{Pocatello}
  \state{Idaho}
  \country{USA}
}
\author{Amiangshu Bosu}

\email{amiangshu.bosu@wayne.edu}
\affiliation{%
  \institution{Wayne State University}
  \city{Detroit}
  \state{Michigan}
  \country{USA}
}

\renewcommand{\shortauthors}{Sultana \em{et} al.}

\input{Sections/abstract}

\ccsdesc[500]{Software and its engineering~Collaboration in software development}
\ccsdesc[500]{Computing methodologies~Supervised learning}
\ccsdesc[500]{Software and its engineering~Integrated and visual development environments}

\keywords{misogyny, sexism, discrimination, hate speech, pejorative}


\maketitle

\definecolor{codegreen}{rgb}{0,0.6,0}
\definecolor{codegray}{rgb}{0.5,0.5,0.5}
\definecolor{codepurple}{rgb}{0.58,0,0.82}
\definecolor{backcolour}{rgb}{0.95,0.95,0.92}

\lstdefinestyle{mystyle}{
    backgroundcolor=\color{backcolour},   
    commentstyle=\color{codegreen},
    keywordstyle=\color{magenta},
    numberstyle=\tiny\color{codegray},
    stringstyle=\color{codepurple},
    basicstyle=\ttfamily\footnotesize,
    breakatwhitespace=false,         
    breaklines=true,                 
    captionpos=b,                    
    keepspaces=true,                 
    numbers=left,                    
    numbersep=5pt,                  
    showspaces=false,                
    showstringspaces=false,
    showtabs=false,                  
    tabsize=2
}

\lstset{style=mystyle}

\newcommand{\surveyquote}[1]{\begin{addmargin}[1em]
{0em}\emph{#1 }\end{addmargin} \vspace{2pt}}

\newcommand{\etal}{{\emph{et} al. }}

\definecolor{background_gray}{gray}{0.85}
\definecolor{background_green}{rgb}{7,163,90}

\newcommand{\bestvalue}[1]{\cellcolor{background_gray}\textbf{#1}}

\newcommand{\greenshade}[1]{\cellcolor{background_green}\textbf{#1}}
\definecolor{MidnightBlue}{rgb}{0.1, 0.1, 0.44}

\newcommand{\code}[1]{{\tt #1}}

\newcommand{\newkeyword}[1]{ \textcolor{blue}{ #1}}

\newcommand{\significant}[1]{ \cellcolor{gray!25} #1}

\input{Sections/introduction}

\input{Sections/related-works}

\input{Sections/methods}

\input{Sections/tool-design}

\input{Sections/results}

\input{Sections/discussion.tex}

\input{Sections/threats.tex}

\input{Sections/conclusion}
\input{Sections/Data_availability}

\bibliographystyle{ACM-Reference-Format}  
\bibliography{bibliography} 

\end{document}

%% file: Sections/abstract.tex
\begin{abstract}
In an industry dominated by straight men, many developers representing other gender identities and sexual orientations often encounter hateful or discriminatory messages. Such communications pose barriers to participation for women and LGBTQ+ persons. Due to sheer volume, manual inspection of all communications for discriminatory communication is infeasible for a large-scale Free Open-Source Software (FLOSS) community. To address this challenge, this study proposes an automated mechanism to identify Sexual Orientation and Gender Identity Discriminatory (SGID) texts in software developers' communications. On this goal, we trained and evaluated SGID4SE (Sexual orientation and Gender Identity  Discriminatory text identification for (4)  Software Engineering texts), a supervised learning-based tool. SGID4SE incorporates six preprocessing steps and ten state-of-the-art algorithms. SGID4SE employs six distinct strategies to enhance the performance of the minority class. We empirically evaluated each strategy and identified an optimum configuration for each algorithm. In our ten-fold cross-validation-based evaluations, a BERT-based model achieves the best performance with 85.9\% precision, 80.0\% recall, and 82.9\% F1-score for the SGID class. This model achieves 95.7\% accuracy and a Matthews Correlation Coefficient of 80.4\%. Our dataset and tool establish a foundation for further research in this direction.

\end{abstract}

%% file: Sections/introduction.tex
{\small \textcolor{red}{ Warning: This paper contains examples of language that some
people may find offensive or upsetting.} }
\section{Introduction}
\label{sec:sec-introduction}
According to the 2023 Stack Overflow developer survey~\cite{so-survey-2023}, only 5.1\% of the professional developers around the world identify as women compared to 91.8\% identifying as men. In an industry dominated by straight men, many software developers harbor sexist, misogynist, and anti-LGBTQ+ beliefs and attitudes, which may vary from subtle to highly overt. Prior research found a presence of sexism and misogyny among various computing organizations~\cite{hanton2015lack,trinkenreich2022empirical,young2015understanding}. For example, Polly, a software engineer from the United Kingdom, shared the worst feedback she received in a code review: \textcolor{teal}{\textit{``I don’t care, I only hired you because you wore a skirt in your interview''}}~\cite{polly-exp}. A 2015 survey titled “Elephant in the Valley'' shows 84\% of women working in Silicon Valley had been called ``too aggressive'' by their men colleagues~\cite{elephant}.  Due to widespread sexist/misogynistic culture and biases against women, 45\% women in computing switch careers within ten years, and that attrition rate is more than twice as high for women than it is for men~\cite{ashcraft2016women}. 
Discriminating attitudes towards  LGBTQ+ persons are also common, as they often are victims of disparaging comments or bullying~\cite{garcia2019better,sultana2021identifying}. These issues are causing attrition of valuable human resources from the computing industry, despite these jobs being in high demand.

\underline{\textbf{S}}exual orientation and \underline{\textbf{G}}ender \underline{\textbf{I}}dentity based \underline{\textbf{D}}iscrimination (SGID)  have been found among many Free and Libre Open source (FLOSS) communities~\cite{paul-SANER-2019,hcissInsult}, as women often encounter colleagues who perceive them as technologically less proficient, are assigned menial tasks, and are subject to sexist/misogynistic insults~\cite{elephant,didio1997}. While women are harassed for mistakes or lack of knowledge, their male colleagues get encouragement for learning from mistakes~\cite{singh2021motivated}. As one woman shared her experience, \textcolor{teal}{\textit{``Oh, she’s a woman. She doesn’t know how to code. That’s why she did something wrong.''}}~\cite{singh2021motivated}. Not only women but also people identifying as LGBTQ+ encounter negative experiences as their identities are used in derogatory ways.  For example, a developer criticized another person's project as \textcolor{teal}{\textit{``This is too gay to be true.  I'm sorry, this is way too gay, plz delete''}}.
Discriminatory comments such as these not only demotivate the participation of women and LGBTQ+ persons but also negatively influence efforts to promote diversity, equity, and inclusion (DEI) among FLOSS projects. Therefore, to promote inclusive computing organizations, it is crucial to combat such SGID comments. 
Although many FLOSS organizations have Codes of Conduct (CoC) to discourage such interactions, these are rarely enforced, as victims are often afraid to report CoC violations due to fears of repercussions~\cite{toxic-blog}. Furthermore, manually checking all interactions is infeasible for project administrators, as large-scale FLOSS communities, such as OpenStack, Wikimedia, Qt, and Apache, regularly generate enormous amounts of text-based communications through various mediums, including code reviews, issue discussions, code commits, and mailing lists. An automated mechanism to flag SGID interactions can assist in two ways. First, it can help project administrators intervene and possibly remove such content. Second, it can also educate people who may not realize that their jokes or remarks are insulting, making many minority groups feel unwelcome. Therefore, the primary objective of this study is \textit{to develop an automated mechanism to identify    Sexual orientation and Gender identity Discriminatory (SGID) texts from software developers' communications.}

Although several recent studies have focused on automated identification of sexist and misogynistic communications, those are limited to Twitter  posts~\cite{jha-mamidi-2017-compliment,problem_of_identification,rodriguez2020automatic}, Reddit discussions~\cite{Farrell,hcissInsult}, and YouTube comments~\cite{shila_1}. 
However, no prior studies have focused on identifying such texts from software developers' communications. A customized SGID tool for the SE is necessary for two reasons. First,   some texts may not be considered sexist in a non-SE context. For example, \textcolor{teal}{\textit{``Documentation! Is there any lady to add documentation?''}} -- without knowing that writing software documentation is considered by many as a menial task, a non-SE classifier is less likely to predict this text as sexist. Second, as prior studies have shown the poor performance of off-the-shelf natural language processing (NLP) tools on Software Engineering (SE) communications~\cite{sarker_1,jongeling2017negative}, off-the-shelf tools are unlikely to achieve reliable performances on a SE dataset. 
Unfortunately, to the best of our knowledge, no such SE domain-specific SGID identification tool or labeled dataset currently exists. To fill this gap, we developed a rubric by conducting a systematic literature review on prior studies that aimed to identify misogynistic texts~\cite{misogyny_rubric}. This study has been published in ESEM 2021. Subsequently, we modified the rubric to cover derogatory texts toward women and LGBTQ+ people. There are a total of 13 categories, of which 12 belong to the SGID group. To address dataset unbalancing, we adopted a keyword-based selection method used in building prior NLP datasets~\cite{waseem-etal-2017-understanding,kumar-etal-2018-benchmarking,anzovino2018automatic}. Specifically, we employed a systematic approach to curate a set of 252 keywords organized into 12 categories. 
After searching the GHTorrent export~\cite{Gousi13} and GitHub search API, we identified 225,117 unique {pull request} comments, including these keywords. After excluding non-English comments using fastText~\cite{bojanowski2017enriching}, we were left with a total of 193,056  comments. As this dataset was still too large for manual labeling, we employed a stratified sampling strategy, as used in prior NLP studies~\cite{sarker_1,sarker2022automated}, to identify 11,007 comments. Each of the selected comments was independently labeled and categorized by two raters with `substantial agreement' in binary (Cohen's $\kappa$~\cite{cohen1960coefficient} =0.658 ) and an `acceptable agreement' in multiclass categorization ( Krippendorff's $\alpha$~\cite{krippendorff2011computing} =0.691). We resolved conflicting labels through mutual discussions. After this step, we identified 1,422 ($\approx$13.6\%) SGID comments belonging to one of the 12 SGID categories. 

Using this dataset, we trained and evaluated SGID4SE (\underline{S}exual orientation and \underline{G}ender \underline{I}dentity based  \underline{D}iscrimination identification for (\underline{4})  \underline{S}oftware \underline{E}ngineering texts), as a supervised learning based SGID detection tool. SGID4SE incorporates six preprocessing steps and ten state-of-the-art algorithms.  We empirically evaluated each strategy and identified an optimum configuration for each algorithm.  In our ten-fold cross-validation-based evaluations, a transformer-based model using the BERT-base encoding~\cite{devlin2018bert} boosts the performance with 85.9\% precision, 80.0\% recall, and 82.9\% F1-score for the SGID class. This model achieves 95.7\% accuracy and 80.4\% Matthews Correlation Coefficient~\cite{chicco2020advantages}. 
 Our post hoc analyses also identify several lessons that can be useful to develop future SE domain-specific SGID tools. 
The primary contributions of this research include:

\begin{itemize}
\itemsep0pt
    \item A classification rubric to manually label SGID texts.
    \item The first labeled SGID dataset from the SE domain.
    \item SGID4SE, an automated SGID detection tool for the SE domain.
    \item Empirical evaluation of optimum configuration for each of the ten algorithms.
    \item A baseline to improve on and a set of lessons for developing future SE domain-specific SGID tools.
    \item We release SGID4SE, a labeled dataset, and evaluation results in the replication package on GitHub at: \url{https://github.com/WSU-SEAL/SGID4SE}
\end{itemize}

\textbf{Notes:} 
   A subset of the authors of this paper previously introduced a rubric for identifying sexist and misogynistic content at ESEM 2021~\cite{sultana2021identifying}. That work focused on categorizing misogynistic remarks, sexist jokes, and speech-based sexist or misogynistic content. While our current study builds upon the rubric proposed in the ESEM 2021 paper, it differs in two key ways. First, we have refined and extended the original rubric by incorporating insights from additional research on the identification of misogynistic content. The details of our rubric development process are outlined in Section~\ref{sec:research-method}. Second, we have implemented an automated tool for detecting such content following the rubric. A preliminary version of this tool was presented in the Student Research Competition at ASE 2022~\cite{sultana2022identifying}. Upon closer examination of the results and datasets, we can find significant differences between these two papers. The SRC paper presents initial findings based on a smaller dataset without any fine-tuning. For instance, the best-performing model in that paper reports 80\% precision, 67.07\% recall, 72.5\% F1 score, and 95.96\% accuracy. In contrast, the current tool enhances performance through several strategies, including adding LGBTQ+ related keywords and samples, pipeline optimization, techniques for addressing dataset imbalance, and error analysis for the best model. As a result, this study reports an improved performance with 85.9\% precision, 80.0\% recall, and 82.9\% F1 score for the SGID class, achieving an accuracy of 95.7\% and a Matthews Correlation Coefficient of 80.4\%.

\textbf{Organization:} The remainder of the paper is organized as follows.
Section~\ref{sec:related-works} discusses related works.
 Section \ref{sec:research-method} describes our methodology for developing a labeled SGID dataset.
 Section~\ref{sec:tool-design} details the design of SGID4SE.
 Sections \ref{sec:results} and \ref{sec:discussion} present evaluation results and discuss the implications of this study, respectively.   
 Section \ref{sec:threats} addresses the limitations, while Section  \ref{conclusion}  concludes the paper.

%% file: Sections/related-works.tex
\section{Related Works}
\label{sec:related-works}
\textbf{Anti-social behavior among software projects.} Numerous research have demonstrated the presence of toxic content in FLOSS communication channels like IRC chat and mailing lists ~\cite{hcissInsult,ferreira2021shut}. Developers working on FLOSS projects have reported insults, attacks, and other forms of toxicity~\cite{Raman2020StressAB}. Miller \etal~\cite{miller2022did} also observed a unique form of toxicity in FLOSS projects that differ from those in other platforms like Reddit or Wikipedia. Such toxic communication comprising entitlements, insult, and arrogance, leads to tension and exhaustion for the developers, and is “likely to make someone leave”~\cite{Raman2020StressAB,miller2022did}. Ferreira \emph{et} al.~\cite{ferreira2021shut} also looked at incivility among contributors in FLOSS projects and discovered that incivility can take many different forms, including bitter frustration, name-calling, mockery, and threats. ``Pushback", a phenomenon where a reviewer blocks the modification request due to a personal conflict, is the result of such uncivil behavior. The Google Jigsaw AI team developed a guidebook~\cite{jigsaw-guideline} for identifying toxic content and the Google perspective API~\cite{perspective-api} for the general domain for automatic toxicity detection.  Sarker \etal~\cite{sarker_1} showed that toxicity detector tools developed for the general domain do not perform well for the particular domain of software developers. The need for automatic toxicity detector tools for software developers prompts the creation of STRUDEL~\cite{Raman2020StressAB} and ToxiCR~\cite{sarker2022automated}.

\vspace{4pt}
\noindent \textbf{Research on sexism and misogyny identification.}
Online misogyny in different platforms, e.g., Twitter~\cite{jha-mamidi-2017-compliment,problem_of_identification, zeinert2021annotating, chiril2021nice}, Reddit~\cite{guest2021expert}, and YouTube~\cite{shila_1} have been subjected to research by quite a few researchers. 
Automated identification of sexist and misogynistic texts can build a healthy environment for women so that they can participate and share their thoughts. On this goal, the Automatic Misogyny Identification (AMI) task hosted at the 2018 IberEval released two labeled datasets of English and Spanish tweets~\cite{fersini2018overview}. They also provided a classification rubric for five types of misogynistic texts: i) Stereotype \& objectification, ii) Dominance, iii) Derailing, iv) Sexual harassment \& threats of Violence, and v) Discredit. This competition resulted in the development and evaluation of several AMI approaches~\cite{anzovino2018automatic,liu2018identification,canos2018misogyny,pamungkas201814,ahluwalia2018detecting,goenaga2018automatic,shushkevich2018classifying,frenda2018exploration,nina2018ami}.
On the one hand, while the tool proposed by  Pamungkas \textit{et} al.~\cite{pamungkas2020misogyny}  achieved the best accuracy of 91\%, it achieved only 36.9\% f1-score in identifying misogynous English texts. On the other hand, the best f1-score of 79\% was achieved by Shushkevich \textit{et} al.~\cite{shushkevich2018classifying} using an ensemble of Naïve Bayes(NB) and Support Vector Machine(SVM) classifiers with an accuracy of 70.6\%. 
Motivated by the IberEval, EVALITA released a labeled AMI dataset of  Italian tweets and hosted a competition to develop tools using that dataset~\cite{fersini2018overview}. This competition was repeated in 2020 using a new dataset~ \cite{basile2020evalita} where Muti \textit{et} al.~\cite{muti2020unibo} achieved the best f1-score of 74.3\% using AlBERTo~\cite{polignano2019alberto}, i.e., a pre-trained BERT model for Italian. 
While datasets and tools for investigating misogyny in the general domain are available and have been explored, misogyny in FLOSS projects has not been studied yet. While investigating profanity and insults in FLOSS projects, Squire \etal~\cite{hcissInsult} found three types of gender-based insults: maternal insult, sexual double entendre jokes, and the use of women relatives to represent unintelligent persons. However, no prior study focused on studying gender-based insults or derogatory content for the specific domain of software developers.

\vspace{4pt}
\noindent \textbf{Relationship between toxicity/incivility and SGID.} Toxicity encompasses a wide range of negative behaviors, with gender discrimination being just one specific type. The Perspective API, developed by Jigsaw and Google, characterizes toxicity as ``A rude, disrespectful, or unreasonable comment that is likely to make you leave a discussion''\footnote{https://developers.perspectiveapi.com/}. Importantly, it does not explicitly address gender issues. As outlined in Sarker \emph{et} al.'s rulebook~\cite{sarker2022automated}, flirtation and identity attacks are included in toxic content. Based on their rules, identity attacks based on gender and sexual orientation and flirtation fall under the definition of toxicity. For example, \textcolor{teal}{\textit {``Why you gay? Why you gay? hmm`` or ``hey pretty girl``} can be identified as toxic and SGID content.} In contrast to Sarker \emph{et} al.'s rulebook, Miller \emph{et} al.~\cite{miller2022did} define toxicity as ``An umbrella term for various antisocial behaviors including
trolling, flaming, hate speech, harassment, arrogance, entitlement,
and cyberbullying.'' Here, the intersection with SGID and toxicity is primarily in the context of harassment related to gender or sexual orientation.
Therefore, SGID contents represent a specific subset of toxicity/incivility. However, not all forms of SGID content are adequately addressed by the toxicity/incivility definitions of Sarker \emph{et} al.~\cite{sarker2022automated}, Miller \emph{et} al.~\cite{miller2022did}, and Ferreira \emph{et} al~\cite{ferreira2021shut}.  For example, prior research has identified stereotyping as a type of misogynistic and SGID content~\cite{anzovino2018automatic, sultana2021identifying}. However, this type of content would not be flagged as toxic according to the standard toxicity definition. For example, \textcolor{teal}{\textit{``What does a blonde do when her computer freezes.......she sticks it in the microwave :P} - this comment expresses stereotyping about women which will not be identified as toxic. But such type of comment should be identified as SGID content.} Therefore, if a generic toxicity detection classifier is used, it may not effectively identify all instances of SGID content in communication excerpts.

%% file: Sections/methods.tex
\section{Dataset to Train and Evaluate Automated SGID Identification Models}
\label{sec:research-method}

We created a large-scale manually labeled SGID dataset with a five-step methodology, as follows: 1) defining SGID, 2) developing a rubric for manual labeling, 3) employing a sampling strategy to create a dataset of SE communications with a higher ratio of SGID texts than randomly selected ones, 4) labeling this dataset using multiple human raters, and 5) retraining existing tools on our dataset. We detail these five steps in the following subsections.

\subsection{Step 1: Research Context Definition  }
Richard Schaefer states, ``Sexism may be defined as an ideology based on the belief that one sex is superior to another"~\cite{schaefer2011sociology}.
This ideology points to biological differences to claim superiority and justify men's dominance over women~\cite{swim1995sexism}. 
While persons from all genders may be the object of sexist attitudes,  women have been the usual victims~\cite{scruton2007palgrave}. 
On the other hand, ``misogyny''  derives from the Ancient Greek word ``mīsoguníā'', which means hatred towards women~\cite{srivastava2017misogyny}. 
Misogyny is often expressed in terms of male dominance, sexual harassment, belittling of women, intimidation, violence against women, and sexual objectification~\cite{kramarae2004routledge,code2002encyclopedia,jane2014back}. 
While earlier literature on sexism and misogyny primarily focus on the binary genders (men vs. women),  prejudice or discrimination towards LGBTQ+\footnote{According to the youth.gov  guidelines (\url{https://youth.gov/youth-topics/lgbt}), LGBTQ+ is an umbrella term that encompasses various sexual orientations such as lesbian, gay, bisexual, transgender, queer or questioning, as well as other diverse gender identities.} persons are not uncommon among software developers~\cite{sultana2021identifying,garcia2019better}.  To include LGBTQ+ persons, this study defines SGID as an umbrella encompassing sexist, misogynistic, and anti-LGBTQ+ expressions. Therefore:

\begin{quote}
{ \textcolor{purple}{\textit{"A text is considered as sexual orientation and gender identity  discriminatory (SGID), if it expresses prejudice or discrimination based on a person's gender, biological sex, gender identity, or sexual orientation.'' }}}\end{quote}

Based on this definition, a straight man can also be a target of SGID. However, our target automated model focuses on SGIDs against minorities (i.e., women and LGBTQ+) since these groups are more likely to be marginalized due to SGIDs.

\begin{table*}
    \centering
    \caption{Rubric to label Gender identity and Sexual orientation Discriminatory (SGID) texts with examples from our dataset}
    \label{tab:rubric}
    \input{Tables/rubric_definition}
\end{table*}

\subsection{Step 2: Classification Rubric}
In the first phase, we focused on developing a binary labeling rubric (i.e., whether a comment is SGID or not). We took the guidelines provided by Guest \etal~\cite{guest2021expert} as our starting point. They provided several inclusion and exclusion criteria to determine when a text should be labeled as misogynous or non-misogynous.
Our adoption of their inclusion criteria includes i) adding prejudice or discrimination against LGBTQ+ persons, ii) adding LGBTQ+ slurs in the list of derogatory terms, and iii) categorizing comments with misogynist or hateful pejoratives towards LGBTQ+ people as SGID.

Although we adopted all of their inclusion criteria by broadening scope, we modified one of their exclusion criteria for two primary reasons. First, while Guest \textit{et} al.'s rubric focuses on identifying misogynous texts, our's focuses on a broader umbrella, including sexism and anti-LGBTQ+ expressions. Second, they did not consider object-directed misogynous pejoratives such as  ``Git is a bitch'' as misogynous. As our rubric is targeted toward professional workplace communication in contrast to theirs being targeted toward everyday Twitter posts, we consider object-directed pejoratives as SGIDs. While some readers may disagree with this change, prior research has shown that words such as `bitch, and `cunt' has strong misogynistic stereotyping roots, and calling someone or some object `bitch' indicates that the target is not conforming to the caller's expected standards similar to a ``malicious, spiteful, or overbearing woman''~\cite{felmlee2020sexist}. Table~\ref{tab:rubric} lists our 12 inclusion criteria with appropriate examples and corresponding categories for SGID texts.

After developing a binary classification rubric for SGID vs. non-SGID comments, we focused on developing an SGID classification scheme to record what type of SGID texts occur more frequently on GitHub. On this goal, we adopted Sultana \textit{et} al.'s sexist and misogynistic text classification rubric developed for the SE domain~\cite{ESEM_sexist}. They proposed three classification schemes for sexist, misogynistic and jokes targeting women. We merged their three schemes into one common umbrella.
We found some categories with different names across these three schemas. We merged such categories into one. For example, we merged one subcategory: `Derailing: reject male responsibility, and attempt to disrupt the conversation to refocus it' with `Victim blaming: blaming the victims for the problems they face' since both express similar notions of misogyny. We also excluded the `mixed bias: gender bias mixed with other types of bias, e.g., religious or regional bias' subcategory since we want to focus only on gender and sexual orientation-related biases. Although Sultana \textit{et} al. did not include a subcategory for maternal insults, we included this subcategory as prior studies have shown texts in IRC chats and emails that involve \textit{`Mom jokes'}~\cite{hcissInsult} or sexist jokes related to woman relatives such as \textit{Grandma test}, \textit{Aunt Tillie test}, and \textit{Girlfriend tests}~\cite{gf_test}.  At this step,  we could map 10 out of our 12 inclusion criteria, each to a different SGID category. To map the remaining two inclusion criteria, we created two additional categories.  The Anti-LGBTQ+ category records prejudice and discrimination against LGBTQ+ persons,  and the `Sexual reference' category records flirtations and references to sexual activities that may be uncomfortable to persons of gender identities. The last column in Table \ref{tab:rubric} lists our mapping from inclusion criteria to SGID categories.
We also include an example for each category taken from our dataset. Among the thirteen categories, twelve belong to the SGID group, and the remaining one forms the Non-SGID group, which includes neutral texts and other inappropriate content that are not discriminatory based on gender or sexual orientation. However, non-SGID texts may still be toxic, racist, or inappropriate for other reasons.

\subsection{Step 3: Dataset Creation}
Although SGID texts exist in software developers' communications, they are not frequent. Even in the general domain, such as Twitter or YouTube comments, a random selection would find a negligible ratio of SGID texts~\cite{zeinert2021annotating}. To overcome such challenges, prior studies selected Tweets based on certain predefined keywords~\cite{anzovino2018automatic} or hashtags~\cite{rodriguez2020automatic}, selected texts from specific Reddit channels~\cite{hcissInsult}, as well as women-oriented blogs and forums~\cite{shila_1}.  Motivated by those examples ~\cite{anzovino2018automatic,waseem-etal-2017-understanding,kumar-etal-2018-benchmarking}, we adopted a keyword-based sampling. 

\subsubsection{Keyword Selection}
 {By analyzing existing SGID datasets and their development methodologies~\cite{anzovino2018automatic,waseem-etal-2017-understanding,zeinert2021annotating,hcissInsult}, we established eleven distinct categories of discussion areas (i.e., the topic that a sentence or paragraph focuses on), all of which can indicate the presence of SGID comments.  The first column of Table~\ref{tab:keyword_list} lists those categories, and the second column provides a brief rationale for why a category may appear in SGID contexts.} After formulating our keyword categories, we looked into prior studies and online lists (listed in Table~\ref{tab:keyword-source}) on sexism, misogyny,  hate speech, pejoratives, and LGBTQ+ terms to identify possible keywords for our categories. We manually inspected each list to identify words fitting one of the eleven discussion areas listed in Table~\ref{tab:keyword_list}. 
We would like to point out that a direct mapping between these discussion areas and our SGID categories listed in Table~\ref{tab:rubric} is not possible, as some of the areas, such as women's roles, women relatives, or general women specific words can appear at various  SGID contexts. For example, `I hate girls' belongs to `Damning' but   `Let's go to score some horny girls' belongs to `Sexual objectification'.
 Table~\ref{tab:keyword-source} also lists the number of keywords taken from each source based on our manual inspections. After aggregating the identified words from the sources and removing duplicates, we identified 215 unique keywords.

\begin{table}[]
    \centering
    \caption{Keyword sources and number of words taken from each source. A word may belong to multiple lists.}
    \begin{tabular}{|p{3.5cm}p{3cm}p{5cm}p{0.5cm}|}    
    \hline    \rowcolor[HTML]{9B9B9B} 
       \textbf{Source}  &\textbf{ Keyword type} & \textbf{Rationale} & \textbf{\#}  \\ \hline
      
       Hewitt \textit{et} al.~\cite{problem_of_identification} & Misogyny context & Authors has listed commonly found misogynistic keywords to study online misogyny  & 17  \\\hline
       \rowcolor[HTML]{EFEFEF} 
       SemEval 2019~\cite{hatEval_dataset} & Hate speech against women & In this dataset organizers listed hateful words against women &  21  \\  \hline
       Guest \textit{et} al.~\cite{guest2021expert} & Pejorative terms for women & These terms are explicitly insulting and derogatory, like "slut" or "whore," or implicitly convey negativity or hostility toward women, such as "Stacy" or "Becky."  & 15  \\ 
       \hline    \rowcolor[HTML]{EFEFEF} 
       Baucom, Erin~\cite{baucom2018exploration} & LGBTQ+ terms & Authors listed keywords that has been used  & 14  \\ \hline       
       Kurita \textit{et} al.~\cite{kurita20towards} & Hatred and identity attack & This list contains words related to women body parts and misogynistic pejorative & 100  \\ 
       \hline    \rowcolor[HTML]{EFEFEF} 
        Hatebase~\cite{hatebase} & Toxic and swear words & This is the world’s largest structured
repository of regionalized, multilingual hate speech and  used in prior study~\cite{Farrell} to identify misogyny in online  & 54  \\ \hline
        List of gendered nouns~\cite{gender-nouns}  &  Gender specific roles & We have taken women-specific gendered nouns and roles that might be used to demean or express stereotypes about women & 11 \\
        \hline      
        \rowcolor[HTML]{EFEFEF} 
        Wikipedia & Pejorative terms for women~\cite{women-term-wiki} & This is the list keywords to belittle or derogate women & 19  \\ \hline       
         Wikipedia & LGBTQ+ terms~\cite{lgbt-term-wiki}  & It contains the list of words that are used to refer LGBTQ+ people & 24 \\ \hline
    \end{tabular}
    
    \label{tab:keyword-source}
    \vspace{-14pt}
\end{table}

\begin{table*}
    \centering
    \caption{List of misogynistic keywords and their groups. Words in \newkeyword{blue} were added during our keyword expansion phase}
    \vspace{-10pt}
    \input{Tables/keyword_list}
    \label{tab:keyword_list}    
    \vspace{-10pt}
\end{table*}

\subsubsection{Keyword expansion}

To identify potentially missing keywords, we loaded the GHTorrent export from March 2021~\cite{Gousi13} in a local MySQL. We queried the \code{pull\_request\_comments} table for all the comments with at least one of our keywords. 
Among the 56.1 million pull request comments from GHTorrent, our search filtered a total of 26,307 comments with our keywords. In the next step, we wrote a Python script using the scikit-learn~\cite{kramer2016scikit} library to compute the frequency of all the words in this corpus. After excluding our initial 215 keywords,  common English stop-words, and words appearing in less than 100 comments, we created a list of 5,316 potentially missing words. Three authors independently went through this list to mark additional words for inclusion. In the next step, they had a joint discussion session to compare their individual lists and argue whether a word should be included. For the conflicting cases, they went with majority votes. At the end of the process, we identified 35 words to add to our keyword list. Keywords listed in \newkeyword{blue} in Table~\ref{tab:keyword_list} are added during our keyword expansion phase.
Although we did not include the `Men roles' group among our initial 11 categories, we added this new group during this phase since words from this group may be used for flirtation or dominance (e.g., ``Who is your daddy?'').

\subsubsection{ Dataset Selection}
 {We create a dataset of pull request (PR) comments on GitHub. 
We selected PR since it is a crucial mechanism to attract contributions from non-members and facilitate newcomers' onboarding among OSS projects~\cite{gousios2014exploratory}. PRs allow contributors to propose changes, which other community members then review. Due to the interpersonal nature of PR interactions and the potential for dissatisfaction due to unfavorable decisions, PR interactions may raise conflicts, frustrations, and incivility~\cite{ferreira2021shut,sarker_1}.}
We identified two possible options to curate a list of SGIDs from publicly available FLOSS projects on GitHub. Initially, we planned to use the GHTorrent MySQL dump~\cite{Gousi13} used for keyword expansion. However, we noticed three shortcomings with this approach. First, GHTorrent MySQL export truncates comments over 255 characters. Therefore, SGID excerpts beyond the 255-character boundary are unavailable in this dataset. Second, GHTorrent export does not include issue comments. Finally, we noticed a relatively smaller number of comments with LGBTQ+ slurs such as `tranny' and `faggot', and misogynistic pejoratives such as `bimbo' and `whore' in this dataset. However, our search using the GitHub search API returned a significantly higher number of such cases than those found in GHTorrent.  This large discrepancy may also be because the most recent GHTorrent export was released over two years earlier.  The second option is to mine directly from GitHub using its search API.  
We decided to use a hybrid approach using both the GitHub search API~\cite{github-search-api} and GHTorrent to mine the comments with each keyword.  First, we mined all the comments with less than 255 characters and our 250 keywords from GHTorrent. We excluded comments of length 255, which may be truncated.  Next, we queried the GitHub search API with our 250 keywords, limiting results to the first 1,000 entries. 
 {While it is possible to add additional filtering criteria, such as date range, to obtain more results beyond the first 1,000 entries from the GitHub search, we did not explore that study since our goal was to create a balanced dataset.  As which categories of SGID a comment belongs to depends on the keywords contained, adding all comments with frequent keywords would create a more unbalanced dataset.} After removing duplicates between the two datasets based on author, text, and timestamps, we had 225,117 unique comments 
  {after excluding bot-generated ones}.

 We also noticed that many comments are written in non-English, as many selected keywords have different meanings in other languages. Moreover, as we cannot comprehend texts primarily written in non-English, we decided to exclude those. We used fastText~\cite{bojanowski2017enriching} to identify the probability each comment is written in English. After excluding all those with a probability of less than 50\% being English, we had 193,056 comments.

However, this dataset is too large for manual labeling. To filter this dataset without discarding the ones more likely to be SGID, we applied a stratified sampling strategy  {proposed by Särndal et al.~\cite{sarndal2003model}. This strategy has also been }used in prior studies~\cite{sarker2022automated,sarker_1} where all the comments are classified using an off-the-shelf classifier that provides a probability of each comment being a positive instance. Then, all the comments with probability above a certain threshold are selected (i.e., Strata 1). To reduce tool bias, certain comments are randomly selected from those with probabilities below the threshold (i.e., Strata 2).

However, finding an off-the-shelf SGID tool was challenging. Despite many recent studies on Automated Misogyny Identification (AMI)~\cite{fersini2018overview}, none of those for English is publicly available. Therefore, we selected a pre-trained BERT-based model proposed by Pamungkas \textit{et} al.~\cite{pamungkas2020misogyny} to re-implement since this model boosted the second highest F1-score during their evaluation, and can be configured to output predictions as probability instead of binary classes opposed to their top performing model. 
During our ten-fold cross-validation-based evaluations using the English AMI IberEval dataset~\cite{fersini2018overview}, this model achieved 81.5\% F1-score and 84.1\% accuracy compared to 83.82\% F1-score and 86.23\% accuracy reported by the authors. We use this implementation (referred to as `the \textit{RefBERT}' hereinafter) to compute the probability that each comment of our dataset is an SGID.  

 {We empirically determined the threshold for our sampling strategy. We wanted to put this threshold as low as possible to ensure we were not missing many SGID samples. We explored various options by lowering the threshold with 0.05 intervals starting at 0.5, the default threshold of RefBERT. With a threshold of 0.2, we obtained a dataset of 5,506 comments that have more than 20\% probability of being an SGID according to the \textit{RefBERT}~\cite{pamungkas2020misogyny}. We noticed that by lowering the threshold further to 0.15, we had to label an additional 3,683 instances manually. As manual labeling is highly time-consuming, we decided against further lowering.  We term this dataset of 5,506 comments as our `Dataset A: High probability with keywords', which comes from the Strata 1. 
We also selected samples from three other strata to mitigate tool biases (i.e., true positive cases potentially missed by RefBert), keyword biases  (i.e., SGID cases with none of our keywords), and missing LGBTQ+ samples. 
To compensate for tool bias, we randomly selected an additional 2,501 comments (99\% confidence interval and 2\% margin of error~\cite{chaokromthong2021sample}) from the remaining comments. We call this set `Dataset B: Low probability with keywords.' 
To account for keyword biases (i.e., SGID cases with none of our keywords), we randomly selected 2,500  comments that do not include any of our 250 keywords. We termed this set as `Dataset C: No keyword.' 
Finally, since we are using an AMI tool for stratified sampling, we noticed a lack of samples with LGBTQ+ words. Although our search found 6,254 comments with LGBTQ+ words, most are bot-generated spam. After filtering out the bot comments, we randomly selected 500 comments from the remaining ones.  We call this set `Dataset D: LGBTQ+ words.' 
Therefore, we selected a total of 11,007 unique comments for our dataset.}

\begin{figure}
	\centering  \includegraphics[width=.95\linewidth]{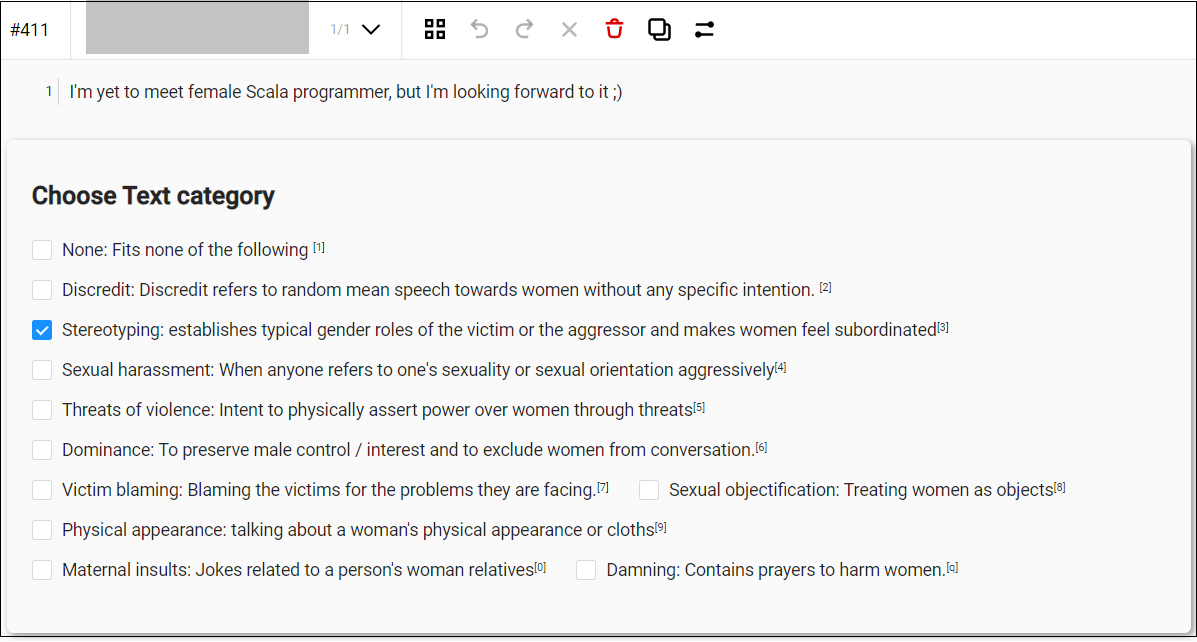}
	\caption{Interface of Label Studio for labeling one entry from the dataset}
	\label{fig:label_studio_ss}	
	\vspace{-12pt}
\end{figure}

\subsection{Step 4: Manual Labeling}
To facilitate our manual labeling process, we installed an instance of \emph{Label Studio}, the Open Source Data Labeling Tool~\footnote{\url{https://labelstud.io/}} in a lab server. To assist with labeling, we configured the labeling interface to include a short definition for each category as listed in Table~\ref{tab:rubric}. The interface shows the raters one comment at a time and allows them to resume from the previous session. A rater could assign multiple categories to a comment.

Figure \ref{fig:label_studio_ss} shows an example from our labeling session. 
Three of the authors worked as raters in this stage. First, they sat together to discuss the inclusion/exclusion criteria and build an agreed-upon understanding. Then, they labeled the top 100 entries (i.e., according to RefBERT probability) from Dataset A through discussions to further validate their understanding. 
We created separate projects with the same dataset for different annotators to avoid biases in the labeling process. We divided the labeling tasks to ensure that at least two of the raters independently labeled each comment. We labeled the dataset in three iterations. First, the raters independently labeled 1,000 entries from Dataset A.  {We noticed still a high number of conflicts with a Krippendorff’s \emph{$\alpha$ = 0.232} indicating a  {`Low agreement'}. They had another discussion session to resolve these conflicts and clarify misconceptions. While resolving conflicts after the first iteration, we found a few cases where comments were mislabeled due to a misunderstanding of the meaning of the texts or context. This session helped build a further improved understanding. We also noticed that binary labeling (i.e., whether a comment fits SGID or not) did not cause many conflicts; rather, which category or categories a comment fits was the primary reason.
After that, they labeled the remaining 4,406 comments from Dataset A. This iteration achieved Krippendorff’s \emph{$\alpha$=0.782} (`Acceptable agreement'). Again, we resolved the conflicts through mutual discussions. The third iteration had the rates independently label the remaining 5,501 instances (i.e., Datasets B,  C, and D) and later held discussions to resolve conflicts. We achieved Krippendorff’s \emph{$\alpha$ = 0.832} (`Satisfactory agreement') during this iteration. Overall, we achieved  Cohen's \emph{$\kappa$ = 0.658}  for binary labeling (i.e., SGID and non-SGID) and  Krippendorff’s \emph{$\alpha$ = 0.691}  for the 13-class labeling.} 
 
 Due to the substantial size of our dataset, which consists of 11,007 instances extracted from pull request messages, conducting a thorough examination of the contexts for each entry would be a time-intensive process. Additionally, it is important to note that this study's primary goal is not an empirical investigation, and therefore, we focused on particular comments rather than delving into the specific contexts of those. It is worth recognizing that not all entries in the SGID class may exhibit discriminatory characteristics when the context is considered.

 At the end of this labeling process, we identified a total of 1,422 ($\approx$ 12.9\%) SGID instances and 9,585 non-SGID instances. 
Table~\ref{tab:dataset} shows the distribution of the final labels among the four datasets.  Since some of the SGID comments belong to multiple categories, the sum of individual category counts is higher than the total number of SGID instances.
 The highest number of SGID entries from our dataset fit into \textit{Sexual Objectification (SO)} category with 660 ($\approx$6\%) instances. \textit{Anti-LGBTQ+} class ranks second  with 339 ($\approx$3\%) cases. Next three classes based on number of instances are:  \textit{Maternal Insult} with 191 ($\approx$1.73\%), \textit{Discredit} with 130  ($\approx$1.2\%),   and  \textit{Appearance reference} with 81 ($\approx$0.7\%). We found only one instance for \textit{Victim blaming}, making it the least common type of SGID in our dataset. However, we should be cautious to take this result as an accurate empirical distribution since our sampling method has limitations (i.e., 255 limit in GHTorrent and 1,000 results from GitHub search API). 
 Guest \textit{et} al.'s~\cite{guest2021expert} dataset of 6,567 English Reddit posts includes 10.6\% misogynous instances, whereas ours includes 12.9\% SGID instances. 
 
\begin{table}
    \centering
    \caption{Distribution of different SGID and non-SGID classes among our labeled datasets }
    \input{Tables/dataset-distribution}
   \label{tab:dataset}

\end{table}

\subsection{Step 5: Evaluation of Existing Tools}
 {First}, we evaluated the performance of RefBERT~\cite{pamungkas2020misogyny} trained on the IberEval dataset to check the performance of an off-the-shelf model (i.e., trained on IberEval but evaluated on our dataset). Second, we retrained RefBERT~\cite{pamungkas2020misogyny} using our dataset to validate the need for a customized pipeline. Third, using our dataset, we retrained four closely related SE domain-specific NLP tools. Our selection of tools includes: 
\begin{enumerate}
    \item ToxiCR~\cite{sarker2022automated} --is trained to identify toxic code reviews.
    \item  STRUDEL tool~\cite{Raman2020StressAB} --is trained to identify toxicity from issue comments.
    \item  { Ferreira \textit{et al.}~\cite{ferreira2024incivility} --is trained to identify incivility from issue comments.}
\end{enumerate}

  Unsurprisingly,  off-the-shelf RefBERT~\cite{pamungkas2020misogyny} performs poorly with only 0.46 precision, 0.342 recall, and 0.392 F1-score for the SGID class. These results validate the need to develop a domain-specific tool considering a domain-specific labeled dataset. After retraining with our dataset, RefBERT~\cite{pamungkas2020misogyny} shows significant performance improvements in precision (0.801), recall (0.693), and  F1-score (0.743). 
  Although retrained STRUDEL performs better than the off-the-shelf RefBERT~\cite{pamungkas2020misogyny}, it fails to outperform RefBERT-retrained. This result may be due to the agnostic TFIDF vectorizer and the SVM algorithm used by STRUDEL, which tend to perform worse than transformers.
  Finally, retrained ToxiCR performs the best among the four tools with 0.815 precision, 0.703 recall, and 0.753 F1-score for the SGID class. These results suggest that SE domain-specific pre-processing helps improve performance for SE domain-specific NLP classifiers since RefBERT~\cite{pamungkas2020misogyny} and ToxiCR differ only in pre-processing steps. 
  While all three models show promising results, their performances for the SGID class are significantly lower than those for the non-SGID class. However, such performance degradation for the minority class may not be surprising since our dataset is highly unbalanced, with approximately 1:6.7 ratio between the SGID and non-SGID samples. Since the identification of SGID samples is the primary goal of this study, we aim to build a custom tool that focuses specifically on improving the performance of the SGID class. Since ToxiCR-retrain provides the best performance, we consider it the baseline for improving our proposed tool.

\begin{table*}[t]
    \caption{Performance of the existing tools on our labeled dataset}
    \label{tab:base}
    \centering
    \input{Tables/no-oversample}
\end{table*}

%% file: Tables/rubric_definition.tex

\resizebox{\textwidth}{!}{
\begin{tabular}{p{6cm}p{6.2cm}p{2.3cm}}
\rowcolor[HTML]{9B9B9B} 
Inclusion criteria                        & Example     & Category     \\
\rowcolor[HTML]{EFEFEF} 
Implies random mean speech to demean women/LGBTQ+ people without   any particular intention                                                   &    \textit{Bimbos can't get them, is this intentional or a bug?} & Discredit    
\\ 
                                                                                 
Derogates women or LGBTQ+ people based on traditional gender roles, personal characteristics,   physical or sexual limitations and to make them  feel  subordinated &   \textit{What does a blonde do when her computer freezes.......she sticks it in the microwave :P} & Stereotyping        \\                                                                             
\rowcolor[HTML]{EFEFEF} 

Refers to or jokes about one's sexuality or sexual orientation in an aggressive manner                                        & \textit{Why you gay? Why you gay? hmm}   & Sexual harassment  \\

Intends to assert power over women or LGBTQ+ persons through threats to intimidate or silence them           & ``BITCH I WILL KILL YOU" & Threats of violence                                                                    \\ 
\rowcolor[HTML]{EFEFEF} 
Intends to control women or LGBTQ+ people to exclude them from the conversation                                 &  \textit{.. I'd love to be able to mute it when the drama queens start using Discord as a soapbox to bypass /ignore.}   & Dominance                                                                                                    \\
Blames the person who encounters problems for  aggression generated by others                                              & ``She is too sensitive, she did not get the joke about women"  & Victim blaming  
 \\
\rowcolor[HTML]{EFEFEF} 
Refers to women or LGBTQ+ people as objects by creating sexual imagery of body parts or themselves                                                 & \textit{Would you like to get some tickets? With horny girls?} & Sexual objectification \\                                                                                                       
Discusses women's or LGBTQ+ people's physical appearance or  clothes                                                                        & \textit{Please Update README.md Blondie is my girl}   & Appearance  reference                                                                                               \\ 
\rowcolor[HTML]{EFEFEF} 
Insults or jokes directed towards a person's women  relatives                                                                       &     \textit{This is cheating harder than your mom does.} & Maternal insults  
\\ 

Expresses ill wish or hatred  towards women or LGBTQ+ persons                                                                       &      \textit{I hate gays, cuz they are really gay}     & Damning                                                                                                                   \\ 
\rowcolor[HTML]{EFEFEF} 

Demeans or insults LGBTQ+  persons or groups  using LGBTQ+ slurs & \textit{ A faggot wrote the source code.} & Anti-LGBTQ+ \\ 

Not directed to women or LGBTQ+ people but mentions uncomfortable  references about gender or sex & \textit{I'm having trouble finding the balls! I know how to ligma but not how to suck} & Sexual references \\
\rowcolor[HTML]{EFEFEF} 

Fits none of the categories &                                                                                                           If \textit{I'd only read the whole source that I cited. Grammar Girl agrees with you. Apologies. Leave it as-is.}         & Non-SGID              \\    
\end{tabular}       

}

%% file: Tables/keyword_list.tex
\resizebox{\linewidth}{!}{
\begin{tabular}{p{1.8cm}p{5.5cm}p{8cm}}
\rowcolor[HTML]{9B9B9B} 
\textbf{Category (\# of keywords)} & \textbf{Rationale} & \textbf{Keywords} \\ 
\rowcolor[HTML]{EFEFEF} 
Pejoratives  (81) & Derogatory adjectives are often used to belittle women or express hostility~\cite{problem_of_identification,Farrell} & bitch, hoe, hysterical, uptight, slag, skank,  slut, dimwit, whore, hormonal, chic, feminazi, chick,  smug, cuck, horny,  cocksucker, cougar, crone,  skintern, bimbo, prostitute, harlot, heifer, gigolo, concubine, bawd, moll, floozy, cheater, witch, frump, wench, twat, sissy, mannish, flirty, ladylike, thot, cenobite, menstrual, vixen, kitten,  hag, bossy, nagging, diva, mumsy, frumpy, cunt, feral, simp, fatcel, femcel, shrew,  pickmeisha, foid, nympho, gold digger, promiscuity, puta, roastie, cock tease, milf, phony, mentalcel, psycho, conchuda,  hustler, streetwalker, spinster, shrill, tart, karen, soccer mom, supermom, bastard, what she said, dowry \newkeyword{bitchy}, \newkeyword{rep-whore} \\ 

LGBTQ+ identities and slurs (30) & LGBTQ+ slurs or identities are often used to express hatred or demean this group
& gay, lesbian, bisexual, transgender, queer, homosexual, gaydar, gaymer, gaysian, tommy, lesbo, sapphic, auntie, artiste, punk, batty, bufty, faggot, fag, pansy, dyke, tranny, trannie,   ladyboy, dickgirl, sheman, shemale, transvestite,  he-she, femboy \\ 

\rowcolor[HTML]{EFEFEF} 
Uncomfor-table  reference (34) & Discussion of sexual acts in software development settings may put women or LGBTQ+ persons in an uncomfortable situation and are barriers to promoting inclusive discussions    & boyfriend, penis, masturbate, dick, arse,  asshole, carnal, erotic, genital,  copulate, copulation, coitus, buttock, lovemaking, get laid, orgasm, virgin,  make love, intercourse, menopause, porn, lust, libido, lewd, salacious, butt, smooch, kiss, \newkeyword{naked}, \newkeyword{banging}, \newkeyword{balls}, \newkeyword{cum}, \newkeyword{cheating}, \newkeyword{lover}   \\ 

Women  kins (15)  & Mom jokes and citations of female relatives are found in content that expresses gendered insults         & mother, mom, grandma, aunt, girlfriend,  momma, mum, grandmother, granny, sister,  niece, mommy, mummy, daughter, mama     \\ 

\rowcolor[HTML]{EFEFEF} 

Woman's body parts (15) & Women's body parts are often used for sexual objectification.          & pussy, vagina, boob, tits, uterus, clitoris,  clit, hymen, breast, nipple, ovary, areola,  vulva, \newkeyword{waist}, \newkeyword{lip}     \\ 

Women  roles (14)    & Traditional beliefs about gender roles also result in discrimination against women          & wife, bride, actress, princess, waitress, queen, mistress, maid, nurse, housewife, heroin, nun, priestess, \newkeyword{wives}   \\    

\rowcolor[HTML]{EFEFEF} 
General  women (13)   &  Women specific words such as `girl', `women' or `female' may be used to talk negatively or express `stereotyping'          & female, females, feminine, maternal, girl, herself, lady, gal,  woman, women, pregnant, pregnancy, feminist, feminism, \newkeyword{ladies}     \\ 

Flirtatious (25)       & Words typically used in flirtatious contexts may allude to sexual harassment or unwanted sexual attentions                                                  & baby, tigress, doll, darling, honey, candy,  dating, sweetie, cutie, sweetheart, babe, \newkeyword{sugar}, \newkeyword{pretty}, \newkeyword{tootsie}, \newkeyword{romeo}, \newkeyword {marry}, \newkeyword{marriage}, \newkeyword{romantic}, \newkeyword{naughty}, \newkeyword{sexy}, \newkeyword{single}, \newkeyword{cute}, \newkeyword{relationship},  \newkeyword{propose}     \\ 

\rowcolor[HTML]{EFEFEF}  
Physical  appearance (13) & Words alluding to physical characteristics are often used to describe  physical appearance       & blonde, curvy, brunette, fugly, tomboy,  busty, skinny, petite, butterface, blondie, \newkeyword{fluffy}, \newkeyword{fat}, \newkeyword{hot}     \\ 

Sexual threat (9)  & Words that express physical and sexual violence can be used to establish dominance and sexual harassment.             & deflorate, hump, fornicate, molest,  sodomize, rape, deflower,  fornication, fuck     \\ 
\rowcolor[HTML]{EFEFEF} 
 
Cloth (11)   & Clothes that are specific to women  or LGBTQ+ persons  can be used in  for sexual objectification or demeaning           & tampon, panty, panties, skirt, bikini, blouse, dress, lingerie, \newkeyword{bra}, \newkeyword{leggings}, \newkeyword{sleeve}   \\ 

\newkeyword{Men roles (5)}   & Reference to man relatives used for flirtation or dominance           & \newkeyword{ dad, papa, daddy, father, husband }     \\ 

\end{tabular}
}

%% file: Tables/dataset-distribution.tex
\begin{tabular}{|l|l|r|r|r|r|r|r|}
\hhline{~~-----}
 \multicolumn{2}{l|}{} & \multicolumn{5}{c|}{\textbf{Dataset}}\\ \hline
\textbf{Group} & \textbf{Subcategory} & \textbf{A} & \textbf{B} & \textbf{C} & \textbf{D}&  \textbf{Overall} \\ \hhline{-------}

\multirow{11}{*}{\textbf{GSD}}& 
Discredit & 120 & 9 & 0 & 1 & 130 \\ \hhline{~------}
&Stereotyping & 24 & 0 &  0 &  0 & 24\\ \hhline{~------}
& Sexual harassment & 11 & 2 & 0 &10 & 23\\ \hhline{~------}
&Threats of violence & 16 & 0 & 0 & 0 & 16\\ \hhline{~------}
&Dominance & 11 & 0 & 0 & 0  & 11 \\ \hhline{~------} 
&Victim blaming & 1 & 0 & 0 &0 & 1\\ \hhline{~------} 
&Sexual objectification & 605 & 46 & 0 & 9 & 660 \\ \hhline{~------}
& Appearance reference & 76 & 4 &0 &1  & 81 \\ \hhline{~------} 
&Maternal insults & 172 & 9 & 0 & 10  & 191 \\ \hhline{~------}
&Damning & 9 & 0 & 0 & 0 & 9  \\ \hhline{~------}
&Sexual reference & 39 & 0 & 0 & 0 & 39  \\ \hhline{~------}
&Anti-LGBTQ+ & 16 & 1 & 0 & 322  & 339 \\ \hhline{~------}
&Others & 72 & 2 & 0 & 0  & 74 \\ \hhline{~------}
&\textit{Total}& 1,019  & 65  &  0 &  338 & 1,422 \\ \hhline{-------} 
\textbf{Non-GSD} & & 4,487 & 2,436 & 2,500 & 162 & 9,585  \\ \hhline{-------}
\textbf{Total} & & 5,506 & 2,501 & 2,500 & 500 &   11,007 \\ \hhline{-------}
\end{tabular}

%% file: Tables/no-oversample.tex
\resizebox{\textwidth}{!}{    

\begin{tabular}{|l|C{1.5cm}|C{1.8cm}|r|r|r|r|r|r|R{1cm}|r|}
\hline
 \multirow{2}{*}{{\textbf{Model}}}  &  \multirow{2}{*}{\textbf{Vectorizer}} & \multicolumn{3}{c|}{\textbf{Non-SGID}} &  \multicolumn{3}{c|}{\textbf{SGID}} & \multirow{2}{*}{\textbf{$A$}}  & \multirow{2}{*}{\textbf{$MCC$}} \\
 \hhline{~~------~~}
 &  &    \textbf{$P_0$} & \textbf{$R_0$} & $F1_0$ & \textbf{$P_1$} & \textbf{$R_1$} & $F1_1$  & & \\  \hline

  RefBERT (off-the-shelf) & BERT-base &  0.901 & 0.941 & 0.923& 0.460 &0.342 &0.392 & 0.863 & 0.321 \\ \hline
  RefBERT (retrain) &  BERT-base &0.941 & 	0.951 &	0.946 &	0.801 &	0.693 &	0.743 &	0.937 &	0.713  \\ \hline

STRUDEL (retrain) & tfidf & 0.958 &	0.955 &	0.956 &	0.701 &	0.715 &	0.707 	& 0.924 &	0.664 \\ \hline

Ferreira \textit{et al.}~\cite{ferreira2024incivility} (retrain) & Bert-base & 0.939 &	0.966 &	0.951 &	0.610 &	0.555 &	0.552 &	0.913&	0.537\\ \hline
 
 ToxiCR (retrain) & BERT-base & 0.957 &	0.975 &	0.966 &	0.815 &	0.703	& 0.753 &	0.940 &	0.723  \\ \hline

\end{tabular}

}

%% file: Sections/tool-design.tex
\section{SGID4SE Design}
\label{sec:tool-design}
The design of SGID4SE is motivated by prior SE domain-specific tools~\cite{sarker2022automated,ahmed2017senticr} and builds on the ToxiCR framework~\cite{sarker2022automated}. The following subsections describe our dataset pre-processing, word vectorization techniques, an overview of the selected algorithms, and additional features added over the ToxiCR to improve performance for the SGID class.

\subsection{Dataset pre-processing}
\label{sec:data-preprocess}
Since pull requests and issue comments often contain URLs, word contractions, emoji markdowns, and source code snippets, we applied the following pre-processing steps to clean comments before using them for training. The first four steps are the same as ToxiCR~\cite{sarker2022automated}. The fifth step is customized to fit the SGID context, and the final step is specific to SGID4SE.

\begin{enumerate} [leftmargin=*]
   
    \item \emph{URL removal:} Many pull request comments contain URLs that refer to external posts or articles. We used a regular expression matcher to identify and remove the URLs.
    
    \item \emph{Contraction expansion:} Contractions are the short form of one or multiple words. Developers use many contractions when communicating with each other. For example: shouldn't $\longrightarrow$ should not, it's $\longrightarrow$ it is. We replaced  153 common contractions with their expanded forms. 
    
    \item \emph{Special symbol removal:} We implemented a regular expression matcher to identify and remove special symbols (e.g., `\&',  `\$').
    
    \item \emph{Splitting identifiers:} During our labeling, we noticed examples of code snippets alluding to SGID categories (e.g., \code{queen.breastSize}).   We used a regular expression matcher to identify and split identifiers written in \textit{camelCase} or \textit{under\_score} forms. For example: \code{current\_bride} $\longrightarrow$ \textit{current bride}, \code{breastSize} $\longrightarrow$ \textit{breast Size}.
    
    \item \emph{Repetition elimination:} To avoid detection, we found instances of intentionally misspelled words. For example, we found one comment as \textcolor{MidnightBlue}{\textit{``Gaaaaaaaaaaaaaaaaay Make it female version of johnson or something, janess?"}} To identify such cases, we used a pattern-based matcher to identify and replace such cases with their respective intended forms.
    
     \item \emph{Emoji removal:} Pull request comments contains emoji. Some of the emojis (i.e., `:ok\_woman:') include words that can interfere with our classifier. We wrote a regular expression-based matcher to identify and replace emoji markups with a neutral word. 

\end{enumerate}

\subsection{Algorithm Selection}
Based on prior SE studies, we have evaluated three groups of algorithms in SGID4SE. Our selection includes four classical and ensemble (CLE) machine learning, three deep neural-network (DNN) )-based, and three transformer-based algorithms. After excluding ToxiCR's two time-consuming yet low-performing algorithms BiLSTM and GBT, SGID4SE evaluates eight out of the ten ToxiCR selected algorithms~\cite{sarker2022automated}. Additionally, SGID4SE adds two pre-trained transformer-based models, ALBERT~\cite{lan2019albert} and SBERT~\cite{reimers2019sentence}, since those have shown state-of-the-art performances for recent SE domain-specific sentiment analysis tasks~\cite{zhang2020sentiment}.
   
\begin{itemize}

\item \textit{Classic and Ensemble-based (CLE) models: }       SGID4SE uses scikit-learn \cite{pedregosa2011scikit} implementations of the following four CLE algorithms: i) Decision Tree (DT), ii) Logistic Regression (LR), iii) Random Forest (RF), and iv) Support Vector Machine (SVM). We use the Term Frequency - Inverse Document Frequency (TF-IDF) vectorizer for the CLE models. Tf-IDf is a vectorization technique based on the Bag of Words (BOW) model that assesses the relevance of a word to a document within a collection and CLE models only works with Tf-IDf. The Tf-IDf score of a word is then calculated by multiplying its term frequency (Tf) by its inverse document frequency (Idf) as: $TfIdf(w, d) = Tf(w, d) * Idf(w)$. Here, $Tf(w,d)$ represents the term frequency (Tf), which is how often a word appears in a document, and the inverse document frequency (Idf), which measures the significance of the word across all documents. The formula for Tf is the frequency of the word in a document divided by the total number of words in that document. Idf is calculated using the formula: $Idf(w) = \log_e \left( \frac{N}{wN} \right) $. Here \( N \) is the total number of documents and \( wN \) is the number of documents containing the word.

 \item \textit{Deep Neural Network (DNN) Models}:
    Using the TensorFlow library, SGID4SE implements three DNN models:  i) Deep Pyramid CNN (DPCNN)~\cite{johnson2017deep}, ii) Long Short Term Memory (LSTM)~\cite{graves2005framewise}, and iii) Gated Recurrent Unit (GRU)~\cite{elsayed2018deep}. We used pre-trained fastText embedding with these algorithms.  FastText, developed by Facebook's AI team, is an efficient method for generating context-free word embeddings. FastText can handle out of vocabulary words by considering the morphological features of words. It creates a word’s vector by combining vectors of its character substrings. As a result, it outperforms other word vectorization methods like Word2Vec and GloVe in natural language processing tasks, especially when the corpus contains unknown or rare words. Therefore, we have used FastText for DNN models. We chose the architecture of the selected models from existing text classifiers~\cite{kurita20towards,sarker2022automated,johnson2017deep,elsayed2018deep}.
    
 \item \textit{Pre-trained Transformer Models (PTM)}:     
        Using tensorflow\_hub, we used pre-trained BERT encoders. We have used three different BERT models. i) bert\_en\_uncased,  commonly referred as  BERT-base model, which is trained with 2,500 million words from Wikipedia and 800 million words from the book corpus; ii) a Lite BERT aka ALBERT base~\cite{lan2019albert}; and iii) a BERT experts model (SBERT), customized using the Stanford Sentiment Treebank (SST-2) for sentiment analysis tasks. 
      
\end{itemize}    

\subsection{Performance Improvement for the Minority Class}
ToxiCR does not implement any strategy to improve the performance of the minority class to encounter unbalanced training data. As our SGID dataset is more unbalanced than ToxiCR's training dataset, SGID4SE implements the following three mitigation strategies. We did not use undersampling since it can result in loss of information if the training dataset is not very large~\cite{haixiang2017learning}. 

\begin{itemize}

    \item \textit{Higher weight for the minority class:} With this strategy, no samples are duplicated or excluded.  During the training, features belonging to the samples from the minority class get higher weights (usually a fixed multiplier sent as a parameter)  than those from the majority~\cite{byrd2019effect}. While an oversampling strategy increases training time, this strategy does not cause such overhead. However, this strategy cannot be applied to all algorithms. While all the neural network and transformer-based algorithms support this customization, only \emph{RandomForest} from the CLE group supports it.
    
    \item \textit{Random oversampling:} In this strategy, randomly selected instances from the minority class are duplicated until a desired ratio between the two classes (i.e., SGID: non-SGID) is achieved~\cite{zheng2015oversampling,haixiang2017learning}.

     \item \textit{Word replacement based new samples:} In this strategy, we manually grouped the keywords from Table~\ref{tab:keyword_list} into 44 GSID equivalent groups. We consider two words belonging to the same group if replacing one with the other in an SGID comment results in a new SGID comment. While the common strategy is to consider only synonyms to generate new samples based on word replacement, we noticed that two words can be equivalent in SGID contexts even if they are not synonyms. For example, while `mom' and  `grandma' are not synonyms, replacing the former in `This is cheating harder than your mom does.' -- creates another SGID. Therefore, the following  nine words -- ``mother'', ``mom'', ``momma'', ``mommy'', ``mummy'', ``mama'', ``grandma'', "granny", and  ``grandmother'',  belong to the same SGID group. Therefore, the word replacement strategy would create eight new samples if the original one includes the word ``mom". 
      {To create equivalent groups, we started with 12 categories from Table~\ref{tab:keyword_list}. If two words belonging to the same cannot be equivalently replaced by one another in an SGID comment from our dataset, we placed them in two different equivalent groups. In the end, we created 44 GSID equivalent groups from the initial 12.}  Our replication package includes the list of words belonging to each group.
     
     Using this strategy, SGID4SE automatically generates additional  SGID samples from the ones from the training set.   During training, randomly selected SGID instances are added to the training set based on the desired ratio between the two classes. For example, if the training set includes $x$ SGID, $y$ non-SGID, and the desired ratio is 0.5, then $(0.5 * y -x)$ generated samples are randomly selected and added to the training set.

      \item \textit{Mixed sampling:} This sampling strategy is a combination of Random and Word replacement-based ones~\cite{haixiang2017learning}, where half of the additional samples to reach the desired ratio are duplicates, and the remaining half are generated ones.
    
\end{itemize}

\subsection{Optimal Threshold Identification} 
During predictions, DNN-based classifiers output the probability of a sample belonging to a particular class. These output probabilities are converted into binary by comparing them against a pre-defined threshold. Many classifiers, including ToxiCR, use 0.5 as the threshold for such conversion. However, recent results suggest that 0.5 may not be the optimum choice, and a model can achieve improved performance by empirically evaluating this parameter~\cite{sarker-esem-2023,baldini2022your}. SGID4SE implements a feature to automatically identify the threshold value that provides the best performance on the test dataset during  10-fold cross-validations of DNN and PTM models.

\subsection{Word count based features}

We implemented word count-based features suggested by Pamungkas \textit{et} al.~\cite{pamungkas2020misogyny} in SGID4SE. We modified the pre-processing
pipeline to count the number of keywords belonging to seven out of the  12 categories from Table~\ref{tab:keyword_list}. We excluded five categories, as we found that keywords from those do not appear frequently among SGID samples. The selected seven categories are: 1) pejorative, 2) women relatives, 3) LGBTQ+ identities and slurs,  4)  women body parts, 5) women roles, 6) women-specific clothes, and 7) women roles. If enabled,  SGID4SE adds an additional seven dimensions to include these word counts after a comment is vectorized.

%% file: Sections/results.tex
\section{Training and Evaluation}
\label{sec:results}
The following subsections detail our training and evaluation of SGID4SE.

\subsection{Evaluation Configuration}
We have used five metrics for evaluation: i) Precision: the percentage of identified cases that belong to that class, ii) Recall: the ratio of the correctly predicted and total number of cases, iii) F1-score: the harmonic mean of precision and recall, iv) Accuracy: the percentage of cases that a model predicted correctly, and v) MCC: it considers the true positives, false positives, false negatives, and true negatives from the confusion matrix and calculates the correlation between the predicted class and true class. The MCC score ranges from -1 to 1 and is considered a more balanced measure than accuracy and F1-score for evaluating binary classification tasks~\cite{chicco2020advantages}.
In our evaluations, we consider MCC as the most important metric to evaluate these models since it is considered the most balanced measure. In case  of an MCC tie, we consider the F1-score for the SGID class since:
i) Identification of SGID comments is our primary objective, and ii) our dataset is imbalanced with 85\% non-SGID comments.
 We evaluated the CLE models using 10-fold cross-validation, where the dataset was split into ten random subsets. Each of the sets was used exactly once for testing, while the remaining nine sets were used for training the model. We computed the average for all eight (i.e., precision, recall, and F1-scores are computed separately for both classes) metrics over the ten runs.  
 For the DNN and Transformer models, our dataset is split into an 8:1:1 ratio where eight sets are used for training, 1 for validation during training, and the remaining for testing. The validation set helps to optimize the hyper-parameters and prevent the model from over-fitting. We set the number of epochs as 30 during training. The validation set is monitored using \code{EarlyStopping} function for the validation loss measure. We set the  \code{patience} parameter to 4 and \code{restore\_best\_weights=True}. This stops models early after a minimum validation loss is achieved, and four consecutive runs do not achieve further lower loss. Weights for the model with the minimum loss are used for testing. 
 As the model performances are normally distributed, we use paired sample t-tests to check if observed performance differences between two configurations are statistically significant ($p<0.05$). We use the `paired sample t-test' since we initialize the random number generator using the same seed to guarantee that cross-validation runs would get the same train/test partition sequences.


\begin{table}[]
    \centering
    \caption{Performance of the algorithms without any performance optimization in SGID4SE. A shaded background indicates significant improvements over the baseline model's performance, i.e.,  ToxiCR(retrain) for a metric.}
    \input{Tables/baseline_models}

    \label{tab:baseline-models}
\end{table}

\subsection{How does each algorithm perform in its basic configuration?}
\label{sec:baseline-model}
Table \ref{tab:baseline-models} shows the performances of the ten selected models with our dataset without any optimization steps. We have highlighted cells with scores that significantly outperform (i.e., $p<0.05$ according to the result of a t-test) our best baseline model (i.e., retrained ToxiCR shown in Table~\ref{tab:base}). The results suggest that CNN, GRU, and BERT significantly outperform our baseline model in terms of MCC and $F1$ score. BERT achieves the best recall among the ten algorithms while also outperforming ToxiCR-retrain's BERT for five out of the eight metrics. 
Since SGID4SE's basic-configuration BERT differs from ToxiCR-retrain's implementation in terms of two pre-processing steps as described in Section ~\ref{sec:data-preprocess}, we can attribute these improvements to those two steps. 

\vspace{4pt}
 \begin{boxedtext}
\textbf{Key finding 1:} 
\textit{Three of the ten algorithms significantly outperform the established baseline regarding the two key metrics (i.e., the F1-score for the SGID class and MCC). Two additional data pre-processing steps  {(repetition elimination and emoji removal)} implemented in SGID4SE provide significant performance boosts for key metrics.}
\end{boxedtext}

\begin{figure}[t]
	\centering  \includegraphics[clip,trim={0 .2cm 0 0},width=\linewidth]{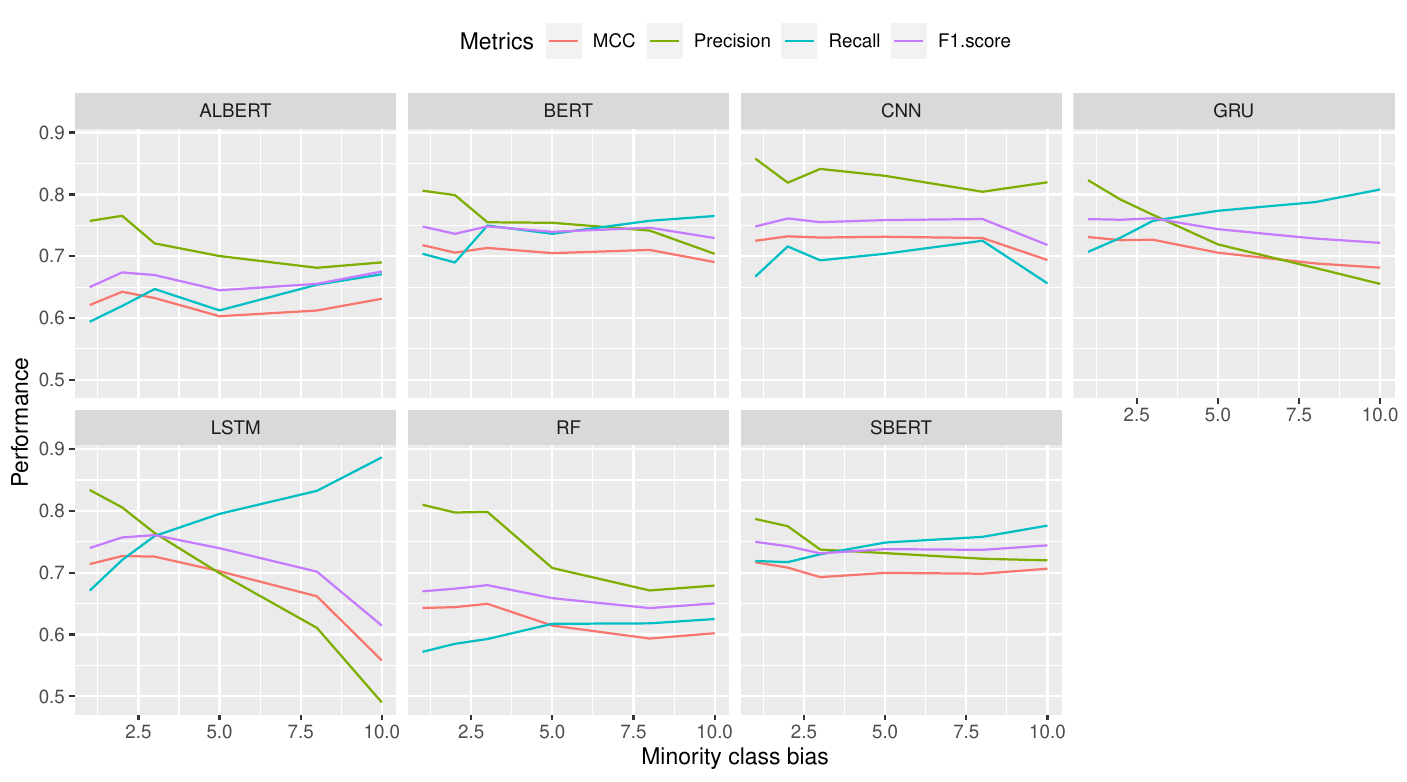}
	\caption{ {The impacts of  minority class weight variations on precision, recall, and F1-score for the SGID class and MCC}}
	\label{fig:class-bias}	
\end{figure}

\subsection{Does assigning higher weights to minority class samples during training provide a performance boost?}  
 Figure~\ref{fig:class-bias} shows the variations of precision, recall, and F1-score for the SGID class and MCC with increasing minority sample weights. As LR, DT, and SVM lack support for class weighting parameters, this analysis was conducted on the remaining seven algorithms. We considered six different class weights for the SGID classes: 2, 3, 5, 8, and 10, with 1 serving as the baseline.
 As expected, precision drops when minority samples get additional weights. 
 For ALBERT, BERT, GRU, LSTM, RF, and SBERT, the recall for the SGID class increases with higher class weights, except for CNN, which experiences a decrease in recall from 66.7\% to 65.6\% with greater class weights. Regarding F1 score, as the class weight increases, the performance of BERT, CNN, LSTM, and SBERT deteriorates, whereas ALBERT exhibits a 2\% improvement, GRU shows a 10\% increase, and RF demonstrates a 5\% increase. 
 However, contrary to our expectations, recall gains are slower compared to the drops in precision. Hence, with this strategy, we see either drops or stagnant MCC and $F1$-score. Only ALBERT exhibits enhanced MCC with a class weight of 10 for the SGID class, while all other models perform {significantly (paired sample t-test, $p < 0.05$)} worse than the baseline class weight. Therefore, this strategy is not viable for achieving higher performance on our dataset.

\vspace{4pt}
 \begin{boxedtext}
\textbf{Key finding 2:} 
\textit{We could not achieve significant {(paired sample t-test)} performance boosts for the key metrics (i.e., MCC ) by assigning higher weights to the SGID samples during training on our dataset.}
\end{boxedtext}

\begin{figure}[t]
	\centering  \includegraphics[clip,trim={0 .2cm 0 0},width=\linewidth]{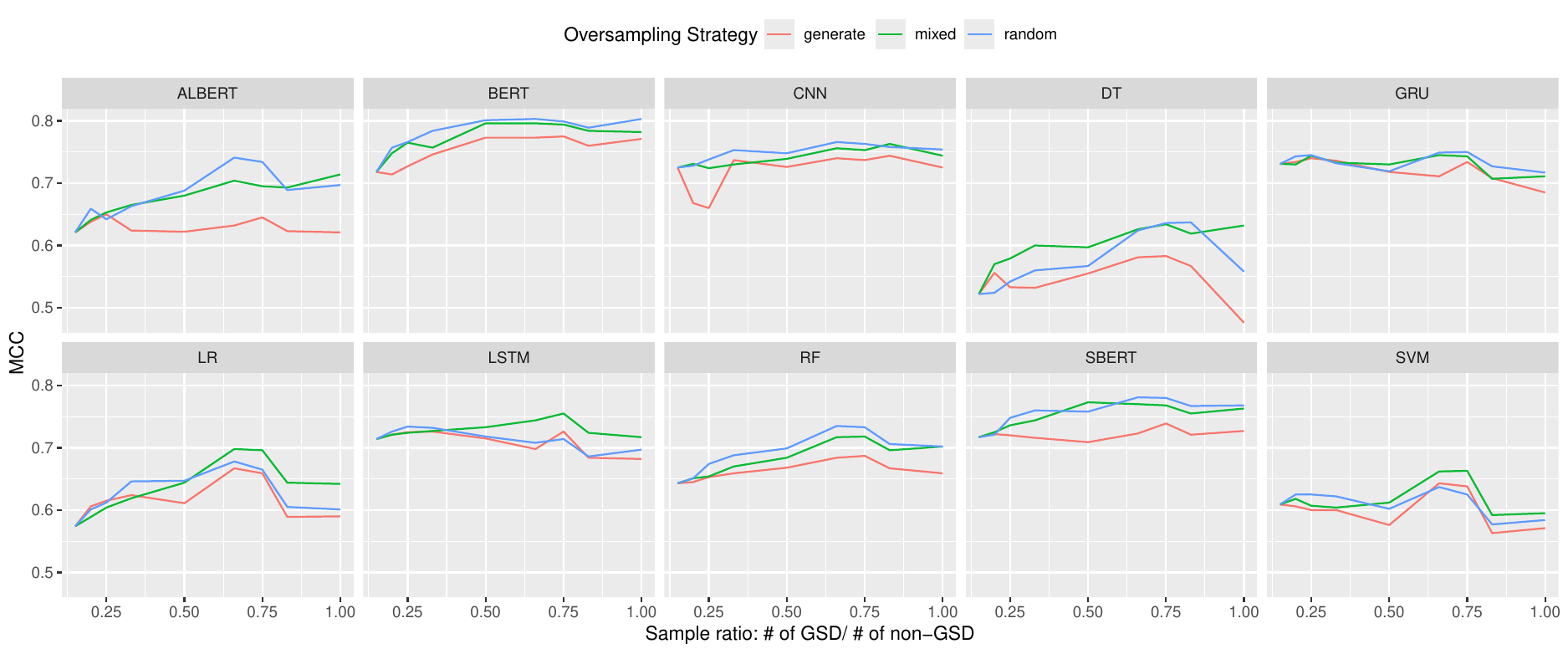}
	\caption{ {MCC change with variation of sampling strategy/ratio}}
	\label{fig:oversampling}	
\end{figure}

\subsection{Which one is the best oversampling ratio/strategy for each algorithm?}

Our models exhibit a bias towards the majority class due to imbalanced SGID and non-SGID ratios, potentially resulting in the neglect of minority class features because of their under-representation. We assessed three oversampling strategies to mitigate such biases: i) random oversampling, ii) generation-based oversampling, and iii) a mixed approach. Figure~\ref{fig:oversampling} illustrates how the MCC of the models vary based on different oversampling methods and ratios, including 0.15, 0.20, 0.25, 0.33, 0.50,  { 0.66, 0.75, 0.83,} and 1. The 0.15:1 ratio serves as the baseline, as that is the original ratio between the two classes in our dataset. On the one hand, we noticed that the random approach provided the best MCC for a particular ratio for seven out of ten algorithms. In contrast, the `mixed' approach ranked second, and the `generate' approach performed the worst.  More importantly, we found that, for each algorithm, the model with optimum oversampling configuration had significantly higher MCC than its baseline performance reported in Table~\ref{sec:baseline-model}. Table~\ref{tab:best-model} shows the best oversampling approach/ratio configuration for the ten algorithms on our dataset.

\vspace{4pt}
 \begin{boxedtext}
\textbf{Key finding 3:} 
\textit{Both random and mixed approaches are viable options to achieve higher MCC scores on our dataset, where random provides the best performance for seven out of the ten models.  {In contrast, the optimum ratio varies between 0.66 to 1 among the algorithms, with 0.66 being the optimum one for five.} }
\end{boxedtext}

\subsection{Does optional word count-based features help achieve higher performances?}

\begin{table}[]
    \centering
    \caption{Performances of the models with word count-based features.  A shaded background indicates significant improvements achieved through word-count features.}
    \input{Tables/wordcount-models}

    \label{tab:wordcount-improvements}
\end{table}
Table~\ref{tab:wordcount-improvements} shows the performance of the models when word-count-based features are enabled. We also performed pair-sampled t-tests to check if performance improvements (if any) are statistically significant and mark such cases with shaded backgrounds in Table~\ref{tab:wordcount-improvements}.
As per the result, word count-based features significantly improved the performances of all four CLE algorithms.

\vspace{4pt}
 \begin{boxedtext}
\textbf{Key finding 4:} 
\textit{Word count-based features significantly improve the performances of all CLE models but do not improve performances for DNN or PTM models on our SGID dataset.}
\end{boxedtext}

\subsection{ How does each algorithm perform with an optimum configuration?}

\begin{table*}[h!]
    \caption{ {Performance of the models with the best combination of word-count features, oversampling method, and the ratio between  SGID and non-SGID} }  
    \label{tab:best-model}
    \centering
    \input{Tables/best-models}

\end{table*}

Table \ref{tab:best-model} shows the performances of the ten models with optimum configurations that we found based on our evaluation of sampling strategy/ratio and word count feature evaluations. 
We have also highlighted the best value for each metric with bold letters. {Furthermore, we have shaded the cells in gray to indicate results that are significantly better than the baseline model ToxiCR (based on t-tests with p-values < 0.05)}
We found BERT emerging as the top-performing model, with five of its measures having the best scores, which include precision for non-SGID class,  F1-scores for both classes, MCC, and accuracy. 
 SBERT remains the second-best model in terms of MCC and F1-score. All three DNN models and two of the three PTM models, excluding ALBERT, outperform ToxiCR-retrain in terms of the two key metrics,  MCC and F1-score for the SGID class.

\vspace{4pt}
 \begin{boxedtext}
\textbf{Key finding 5:} 
\textit{BERT boosts the best performance with 85.9\% precision, 80.0\% recall, 82.8\% F1-score, 95.7\% accuracy, and 80.4\% MCC with random-oversampling. Five neural network-based models outperform the baseline ToxiCR-retrain.}
\end{boxedtext}

\subsection{Do optimal threshold selection improve performance for the best models?}
\begin{figure}[t]
	\centering  \includegraphics[clip,trim={0 .2cm 0 0},width=\linewidth]{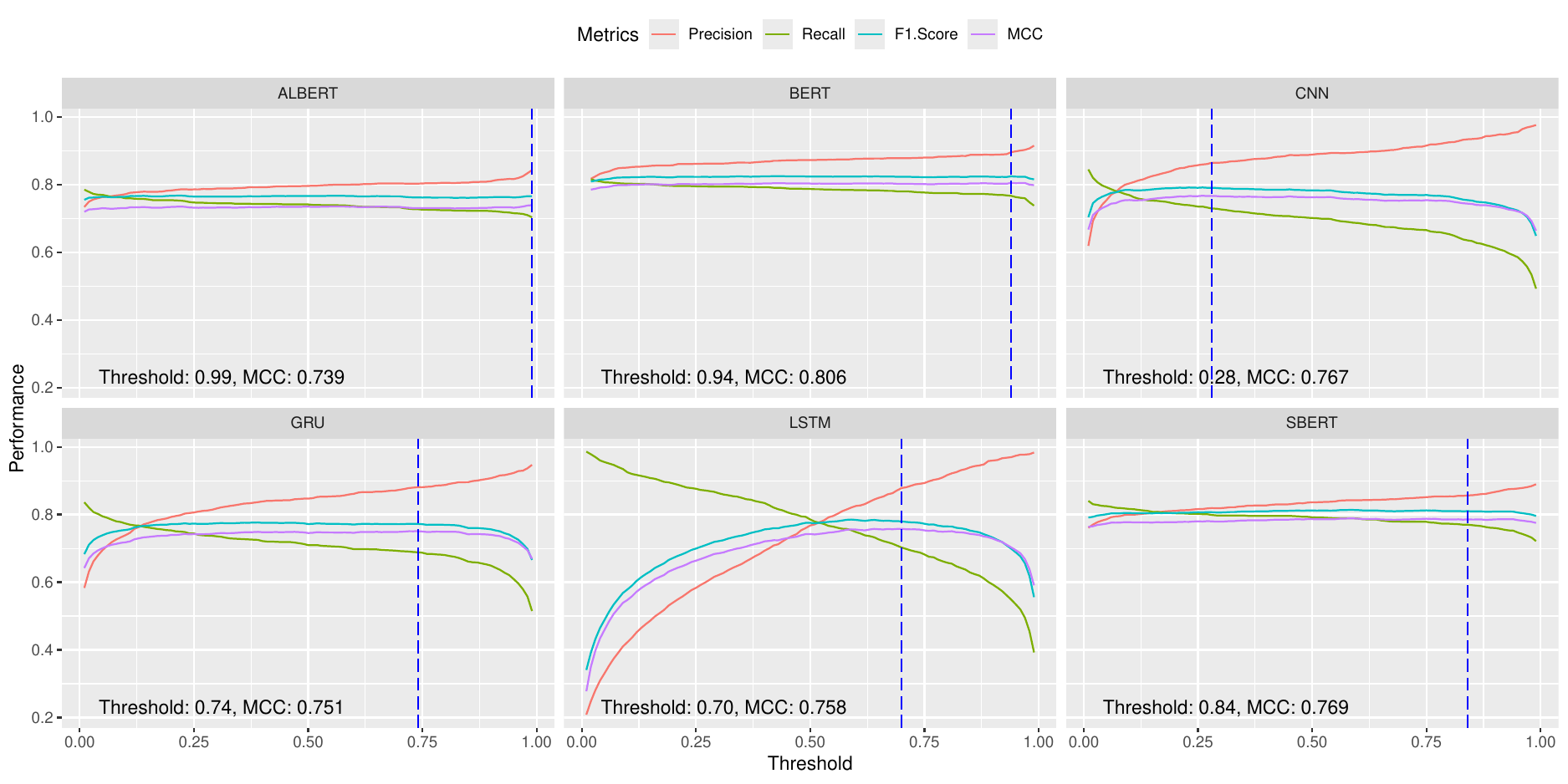}
	\caption{ {Performance improvement with threshold variation. }}
	\label{fig:threshold-selection}	
\end{figure}

\begin{table}[]
    \centering
    \caption{ {Performances of the DNN and PTM models with the threshold to maximize MCC.  A shaded background indicates significant improvements over the model with the default threshold (0.5).}}
    \input{Tables/threshold-improvement}

    \label{tab:threshold-improvements}
\end{table}

Since only DNN and PTM support this parameter, this analysis only applies to the six algorithms belonging to those groups. With the optimum configurations from Table~\ref{tab:best-model}, we varied the threshold from 0.1 to 0.99 and computed ten-fold cross-validation performances for each threshold.
Figure~\ref{fig:threshold-selection} shows variation in performances for the SGID class and MCC score with varying threshold and optimum threshold values. We noticed that the optimum threshold for all six models is higher than the default value (0.5). We also noticed that DNN models are more sensitive to thresholds than the PTMs and encounter wider variations. 
Table~\ref{tab:threshold-improvements} shows the performances of the DNN and PTM models with empirically determined optimum thresholds. We also performed `pair sampled t-tests' to identify statistically significant improvements that we show with shaded backgrounds. Notably, we noticed significant improvements in precision for the SGID class for five of the six models. However, we did not notice significant improvements in the two key metrics, MCC and F1-score of the SGID class. 
While BERT with a 0.98 threshold provides the best MCC, its improvement over the model with a default threshold is not statistically significant. Therefore, we still consider BERT with a 0.5 threshold as the best-performing model and report this model's performance as the best one throughout the paper.

\vspace{4pt}
 \begin{boxedtext}
\textbf{Key finding 6:} 
\textit{While empirically identified optimum threshold improves MCC, F1-score of the SGID class, and MCC, these improvements are not statistically significant. However, optimum thresholds do significantly improve precisions of the SGID class.}
\end{boxedtext}

\subsection{What are the most common misclassifications from the best-performing model?}

\begin{table}[t]
    \caption{Confusion Matrix for our best-performing model (i.e.,  BERT ) } 
    \label{tab:conf_matrix}
    \centering
    \input{Tables/confusion_matrix}
    
\end{table}

To better understand our model, we have analyzed the misclassifications of the best-performing (i.e., BERT with a  0.5 threshold) model. Table~\ref{tab:conf_matrix} shows the confusion matrix of this model, which misclassifies 476 comments out of 11,007 comments, with 190 false positives (FP) and 286 false negatives (FN).  We followed an open coding approach to categorize the misclassified ones. After independently scrutinizing misclassified ones, two authors discussed these errors to create a classification scheme for false positives with four categories. Then, these raters independently tagged each misclassified case with one of those four categories. They compared the assigned tags and resolved conflicts through another discussion. For the False negatives, we looked into our 12 class SGID schema from Table~\ref{tab:rubric}.

\vspace{4pt}
\noindent\textbf{{False positives (FP).}}  

\begin{enumerate}[leftmargin=*]
 \item \emph{Project Discussion (PD):} GitHub hosts many `R-rated' open source games. Pull request discussions for such projects include sexually explicit language as game scenarios. For example,  \textcolor{MidnightBlue}{\textit{``suggestion:  If the butcher didn't cut off the skin, the tits here are more recognizable.''}}, is suggesting to change the message shown. We did not label such cases as SGIDs, but our classifier predicted those as ones. Overall, 33\% of the FPs belong to this category.

 \item \emph{Pre-processing induced errors:}
 We implemented regular expressions to identify adversarial patterns where one or two characters are intentionally misplaced or repeated (e.g., `b00b', and  `gaaay') to avoid getting flagged. While these patterns were successful in fixing most of such cases, those also replaced some of the benign phrases (e.g., `booboo', and `boob') with SGID ones. For instance, \textcolor{MidnightBlue}{\textit{``Can i keep the honey boo boo?''}}. Around 11.8\% of FPs belonged to this category.

 \item \emph{General Error (GE):} 
 The number of SGID instances in our dataset is relatively small.  Therefore, our dataset does not have adequate samples to build a highly precise classifier. As a result, our model misclassifies some comments as SGID if those contain words from the `General women' group (see Table~\ref{tab:keyword_list}). We label such errors as GE, which includes 51.13\% of the FP cases. For example, \textcolor{MidnightBlue}{\textit{``One of my random blonde moments; will fix.''}} is incorrectly classified as an SGID by our model.

 \item \emph{Toxic but not SGID:} Although toxic, some comments do not fit the criteria to get labeled as SGID. For example, \textcolor{MidnightBlue}{\textit{``Please make their yield be five so they don't fucked over by Rng''}} is toxic but not SGID. Almost 4.07\% FPs belong to this group.

\end{enumerate}

\vspace{4pt}
\noindent\textbf{{False Negatives (FN).}} 
We also analyzed the FN cases to identify which SGID categories were more frequently missed by our model. Not only is our dataset imbalanced, but approximately 50\% of the SGID instances belong to the sexual objectification category (See Table~\ref{tab:dataset}). 
We noticed that categories with fewer instances were more likely to be missed. For example, we had only 23 sexual harassment cases, and our model failed to identify 8 out of those 23 with a false negative rate of 34.7\%.  We found 56.15\% false negatives for `Discredit' despite this category having 130 instances and 41.67\% of FNs for `Stereotyping'. Pre-processing steps targeting this category can help train a better SGID classifier using our dataset.
The lowest false negative rate (i.e., 8\%) is seen for the anti-LGBTQ+ class, as most of the instances from this class contain anti-LGBTQ+ slurs. Both `Maternal insults' (3.14\%) and `Sexual objectification' (11\%) had lower false negatives than the overall model, as each of those classes had more than 100 instances, and identifying those cases can be straightforward because of the presence of SGID-related keywords.

\vspace{4pt}
 \begin{boxedtext}
\textbf{Key finding 7:} 
\textit{Presence of keywords associated with gender groups such as `woman', `girl', `female', `females' are more commonly associated with false positives. The best model frequently misses instances belonging to Discredit and Stereotyping. }
\end{boxedtext}

%% file: Tables/baseline_models.tex
\resizebox{\textwidth}{!}{
\begin{tabular}{|l|C{1.5cm}|c|r|c|r|r|r|r|R{1cm}|r|}
\hline
 \multirow{2}{*}{\textbf{Group}} & \multirow{2}{*}{\textbf{Algo}} &  \multirow{2}{*}{\textbf{Vectorizer}}&  \multicolumn{3}{c}{\textbf{Non-SGID}} &  \multicolumn{3}{|c|}{\textbf{SGID}} & \multirow{2}{*}{\textbf{$A$}}  & \multirow{2}{*}{\textbf{$MCC$}} \\
 \cline{4-9}
 & & &    \textbf{$P_0$} & \textbf{$R_0$} & $F1_0$ & \textbf{$P_1$} & \textbf{$R_1$} & $F1_1$  & & \\ \hline

  \multirow{4}{*}{CLE} &  DT & tfidf  & 0.941&	0.956&	0.948	&0.670	&0.598&	0.631&	0.910	&0.581
\\\hhline{~----------}
& LR & tfidf  & 0.924 &	\significant{\textbf{0.997}}	&0.959	& \significant{\textbf{0.957}}&	0.449	&0.610&	0.926&	0.627

\\\hhline{~----------}
&  RF & tfidf  & 0.940	&\significant{0.984}&	0.961&	\significant{0.845}&	0.575&	0.683&	0.931&	0.662
\\\hhline{~----------}
 & SVM & tfidf  & 0.935&	\significant{0.992}&	0.963	&\significant{0.911}&	0.536&	0.674	&0.933&	0.668
\\\hhline{-----------}

 \multirow{4}{*}{DNN} & CNN & fasttext  & 0.954&	\significant{0.983}&	\significant{\textbf{0.968}}&	\significant{0.870}	&0.680&	\significant{0.756}&	\significant{\textbf{0.944}}&	\significant{\textbf{0.736}}
\\\hhline{~----------}
 & GRU & fasttext  & 0.957&	\significant{0.978}&	\significant{0.967}&	\significant{0.829} &	0.703&	\significant{\textbf{0.761}}&	\significant{{0.943}}&	\significant{{0.732}}
\\\hhline{~----------}
& LSTM & fasttext  &  0.948	&\significant{0.985}&	0.966&	\significant{ 0.867}&	0.634&	0.730&	0.940&	0.709
\\\hhline{-----------}

 \multirow{4}{*}{PTM} & ALBERT & bert & 0.945&	0.966&	0.955	&0.743&	0.618&	0.667&	0.921&	0.631
\\\hhline{~----------}
& BERT & bert-base  & \significant{\textbf{0.960}}	&0.973	&0.966&	0.806	&\significant{\textbf{0.722}} &	\significant{0.758}&	\significant{0.941}&	\significant{0.728}
\\\hhline{~----------}
& SBERT & bert  & \significant{0.959}	&0.969&	0.964&	0.776	&\significant{0.721}&	0.745	&0.937&	0.711
\\\hline

\end{tabular}
}

%% file: Tables/wordcount-models.tex

\begin{tabular}{|l|C{1.5cm}|r|c|r|r|r|r|R{1cm}|r|}
\hline
 \multirow{2}{*}{\textbf{Group}} & \multirow{2}{*}{\textbf{Algo}} &   \multicolumn{3}{c}{\textbf{Non-SGID}} &  \multicolumn{3}{|c|}{\textbf{SGID}} & \multirow{2}{*}{\textbf{$A$}}  & \multirow{2}{*}{\textbf{$MCC$}} \\
 \cline{3-8}
 &  &    \textbf{$P_0$} & \textbf{$R_0$} & $F1_0$ & \textbf{$P_1$} & \textbf{$R_1$} & $F1_1$  & & \\ \hline

  \multirow{4}{*}{CLE} &  DT   & \significant{0.946}	&0.955&	\significant{0.951}&	0.677	&\significant{0.634}&	\significant{0.654}&	\significant{0.913}&	\significant{0.606}
\\\hhline{~--------}
& LR   &\significant{0.933}&	0.992&	\significant{0.962}&	0.909&	\significant{0.522}&	\significant{0.663}&	\significant{0.931}&	\significant{0.658}
\\\hhline{~---------}
&  RF   & \significant{0.946} &	\significant{0.985}&	\significant{0.965}&	\significant{0.863}&	\significant{0.623}&	\significant{0.723}&	\significant{0.938}&	\significant{0.701}
\\\hhline{~---------}
 & SVM   & 0.942&	\significant{0.987}&	\significant{0.964}&	\significant{0.872}&	0.588&	\significant{0.702}&	\significant{0.936}&	\significant{0.684}
\\\hhline{----------}

 \multirow{4}{*}{DNN} & CNN   & 0.957&	0.978&	0.967&	0.837&	0.706&	0.762&	0.943&	0.735
\\\hhline{~---------}
 & GRU   & 0.956&	0.978&	{0.967} &	{0.828}	&{0.693} &	{{0.753}}&	{{0.942}} &	{{0.725}}
\\\hhline{~---------}
& LSTM   &  0.950	&0.984&	0.967	&{0.865}	&{0.654}	&{0.743} &	{0.941}	& {0.720}
\\\hhline{----------}

 \multirow{4}{*}{PTM} & ALBERT  & 0.947&	0.973	&0.959&	{0.785} &	{0.627} &	{0.688} &	{0.928} &	{0.659}
\\\hhline{~---------}
& BERT   & 0.958 &	0.970	&0.964	&{0.784} &	{0.711}	& {0.743} &	{0.937}	& {0.710}
\\\hhline{~---------}
& SBERT   & {0.959} &	0.969&	0.964	&{0.777} &	{{0.717}} &	{0.744}	& {0.936} &	{0.710}
\\\hline

\end{tabular}

%% file: Tables/best-models.tex
\resizebox{\textwidth}{!}{
\begin{tabular}{|c|C{0.9cm}|C{2.1cm}|C{2cm}|r|r|r|r|r|r|R{1cm}|r|}
\hline
  \multirow{2}{*}{\textbf{Algo}} &  \multicolumn{3}{c|}{\textbf{Best Configuration}}&  \multicolumn{3}{c}{\textbf{Non-SGID}} &  \multicolumn{3}{|c|}{\textbf{SGID}} & \multirow{2}{*}{\textbf{$A$}}  & \multirow{2}{*}{\textbf{$MCC$}} \\
 \hhline{~---------~}
 & \textbf{Word count} & \textbf{Oversample approach} &\textbf{ \#non-SGID / \# SGID}  &    \textbf{$P_0$} & \textbf{$R_0$} & $F1_0$ & \textbf{$P_1$} & \textbf{$R_1$} & \textbf{$F1_1$}  & & \\  \hline
    
DT & $\checkmark$ & random & 0.83	&\significant{0.957}&	0.944 &	0.951 &	0.656 &	\significant{0.717}  &	0.685 &	0.915 &	0.637
 \\\hline

LR & $\checkmark$ & mixed & 0.66 &	0.954 &	0.973 &	0.963 &	0.792 &	0.681 &	0.732 &	0.935 &	0.698
\\\hline

RF & $\checkmark$ & random & 0.66 &	\significant{0.957} &	0.980 &	0.968 &	0.836 &	\significant{0.703} &	\significant{0.763} &	0.944 &	0.735
\\\hline

SVM & $\checkmark$ & mixed &  0.75 &	0.957 &	0.955 &	0.956 &	0.700 &	0.714 &	0.707 &	0.924 &	0.663
\\\hline

CNN & $X$ & random & 0.66&	\significant{0.957} &	\textbf{0.987} &	\significant{0.972} &	\textbf{0.889} &	0.702 &	\significant{0.783} &	\significant{0.950} &	\significant{0.763}

 \\\hline

GRU & $X$ & random & 0.75 &	\significant{0.959} &	0.980 &	0.970 &	0.851 &	\significant{0.714}	& \significant{0.776} &	0.947 &	0.750

\\\hline

LSTM & $X$ & mixed & 0.75 &	\significant{0.969} &	0.966 &	\significant{0.968} &	0.782 &	\significant{0.794} &	\significant{0.786} &	\significant{0.944} &	\significant{0.755}

\\\hline

ALBERT & $X$ & random & 0.66 & \significant{0.962} &	0.973 &	0.968 &	0.803 &	\significant{0.743} &	0.772 &	0.943& 	0.738
\\\hline

BERT & $X$ & random & 1  &0.971	&0.980	&\significant{\textbf{0.975}}&	0.859&	\significant{0.800	}&\significant{\textbf{0.828}}&	\significant{\textbf{0.957}}&	\significant{0.804}
 \\\hline

SBERT & $X$ & random  &0.66 &	\significant{\textbf{0.972}}&	0.971 &	0.971 &	0.810&	\significant{\textbf{0.811}} &	\significant{0.808} &	0.950&	\significant{0.781}
\\\hline
     
\end{tabular}
}

%% file: Tables/threshold-improvement.tex
\definecolor{dartmouthgreen}{rgb}{0.05, 0.5, 0.06}
\begin{tabular}{|C{1.5cm}|r|r|c|r|r|r|r|R{1cm}|r|}
\hline
  \multirow{2}{*}{\textbf{Algo.}} &  \multirow{2}{*}{\textbf{Threshold}}&  \multicolumn{3}{c}{\textbf{Non-SGID}} &  \multicolumn{3}{|c|}{\textbf{SGID}} & \multirow{2}{*}{\textbf{$A$}}  & \multirow{2}{*}{\textbf{$MCC$}} \\
 \cline{3-8}
 & &     \textbf{$P_0$} & \textbf{$R_0$} & $F1_0$ & \textbf{$P_1$} & \textbf{$R_1$} & $F1_1$  & & \\ \hline

CNN & 0.28  & 0.961 &	0.983 &	0.972 &	0.865 &	\significant{0.730} &	\significant{0.791} &	0.950 &	0.767

\\\hhline{----------}
  GRU & 0.74 &  0.955 &	\significant{0.986} &	\significant{0.970} &	\significant{0.881} &	0.688 &	0.772 &	0.948 &	0.751

\\\hhline{----------}
 LSTM & 0.70  & 0.957	&\significant{0.985}&	\significant{0.971}&	\significant{0.878} &	0.703&	0.780 &	\significant{0.949} &	0.758

\\\hhline{----------}

 ALBERT & 0.99 & 0.957 &	\significant{0.980} &	\significant{0.969} &	\significant{0.843} &	0.704 &	0.766 &	\significant{0.945} &	0.739

\\\hhline{----------}
 BERT & 0.94  &     {0.968} &	 \significant{0.984} &	\significant{0.976} &	\significant{0.883} &	0.782 &	0.828	& 0.958 &	0.807

\\\hhline{----------}
 SBERT & 0.84  &0.966&	\significant{0.981} &	0.973 &	\significant{0.857} &	0.770 &	0.810 &	0.953 &	0.786

\\\hline

\end{tabular}

%% file: Tables/confusion_matrix.tex
\begin{tabular}{l|l|c|c|c}
\multicolumn{2}{c}{}&\multicolumn{2}{c}{\textit{Predicted}}&\\
\cline{3-4}
\multicolumn{2}{c|}{}&\textbf{SGID} & \textbf{Non-SGID}&\multicolumn{1}{c}{}\\
\cline{2-4}
\multirow{2}{*}{\textit{Actual}}& \textbf{SGID} & $1,138$ & $286$ & \\
\cline{2-4}
& \textbf{Non-GSD} & $190$ & $9,393$ & \\
\cline{2-4}

\end{tabular}

%% file: Sections/discussion.tex
\section{Discussion}
\label{sec:discussion}

The following sections discuss key lessons based on our study and provide several directions for future research.

\subsection{Findings:}
The following are the key findings based on this study.

\noindent \textbf{Lesson \#1: SGIDs on GitHub differ from social media.}:
Sexual harassment is one of the most frequent categories of SGID on social media, with more than one-fifth of the cases~\cite{fersini2018overview,basile2020evalita}. However, we noticed less than 1.7\% sexual harassment (23) among our SGID cases.
Among the 1,422 SGID instances, SO ($\approx$46\%) and Anti-LGBTQ+ ($\approx$23.8\%) are the most frequent ones. Prior studies show a dominance of particular categories of misogyny in different languages. A strong representation of \textit{`neosexism'} was found in Danish tweets where discrimination against women is questioned and men are presented as victims~\cite{zeinert2021annotating}.  Mulki \etal~\cite{arab_text} introduced a new category named \textit{`damning'} while labeling misogynistic content in Arabic tweets. Though it is unclear whether language, context, or culture impact the dominance of a particular type of SGID, it requires further investigation. 

Similar to existing lexicon-based data collection studies~\cite{Farrell}, our study may suffer from a lack of completeness because of a limited collection of keywords. From the labeled SGID texts, we can target a few projects where SGID-positive comments were generated and explore the existence of SGIDs in depth. That will assist in mitigating lexicon-based bias and increase the variety of linguistics~\cite{guest2021expert}. However, almost 73\% instances belonging to three groups imply sexual objectification, anti-LGBTQ+ comments, and discredits are the dominant forms of SGIDs on GitHub.  These findings also align with prior FLOSS studies~\cite{singh2021motivated, hcissInsult}.

On the other hand, some of the interactions that may seem non-SGID during general conversations are more likely to be SGIDs during pull request contexts. For example, we noticed maternal insults (aka `mom jokes') as the third most frequent SGID category. These cases often represent women as computer illiterate, nagging, or sexual objects. For example, we found a pull request comment saying, \textcolor{MidnightBlue}{``okay, mom!''}. Upon further investigation, we found that the author complained about a reviewer's insistence on some changes. However, in a general context, this text is not an SGID. But in this context, it implicitly says {``okay, [you are nagging like a] mom!''}. Therefore, in this context, it can be considered an SGID.


\vspace{4pt}
\noindent \textbf{ Lesson\#2: Existing toxicity detector tools do not perform well for SGID content.} We evaluated the performance of STRUDEL and ToxiCR tools on our dataset and found that they did not yield satisfactory results in detecting SGID content {where ToxiCR gives the best precision as 81.5\%}. However, GRU and BERT exhibited superior performance compared to these tools, even without any performance-enhancing strategies. Notably, after fine-tuning with an oversampling method, BERT achieved an impressive 95.7\% accuracy and a precision of 84.1\%. Consequently, our findings underscore the need for a dedicated tool tailored to the task of SGID content identification. 


\vspace{4pt}
\noindent \textbf{Lesson \#3: Application of SGID4SE to combat SGIDs.}  {BERT outperformed all other models with the best performance with 85.9\% precision, 80.0\% recall, 82.8\% F1-score, 95.7\% accuracy, and 80.4\% MCC with random-oversampling technique. Future researchers should consider this model as a baseline for their studies. However, since both the precision and recall of our best model are close to 80\%, it misses approximately one in five SGID cases. While we concur that significant improvement opportunities remain for SE domain-specific SGID models, a project can utilize our model to automatically flag potential SGIDs. A project administrator can review a flagged communication and inspect the context to make a final determination. SGID4SE's 86\% precision indicates that one out of the seven flagged cases is more likely to be a false negative.

\vspace{4pt}
\noindent \textbf{ Lesson \#4: Games-related projects harbor SGIDs due to the target audience.}
The \#gamergate brought attention to sexism and misogyny in the gaming community~\cite{massanari2017gamergate}. We also noticed many comments among gaming projects that include SGID words. However, in most of those cases, we found those as character dialogues. Considering the character quotes, we labeled those as non-SGIDs. However, many such cases appeared as false positives by our classifiers. 

\vspace{4pt}
\noindent \textbf{ Lesson \#5: Regulations are necessary for naming conventions.} We found the usernames of contributors that contain misogynistic keywords. For instance, \textcolor{MidnightBlue}{\textit{hornygranny}}. It should be enforced that users must avoid misogynistic words while creating usernames, variable names, or library names. Another instance, \textcolor{MidnightBlue}{\textit{``Actually, this is still wrong, since `listOfToxinsInThisBitch` is a list of types, the for loop will filter everything out.''}} Using such sexist words in variable names may create discomfort for women developers. Also, many libraries are named with misogynistic keywords. For instance, \textcolor{MidnightBlue}{\textit{``s/thosecunts/helpdesk/g''}}. Not surprisingly, due to the absence of women, misogynistic keywords are used in naming libraries or resources. Therefore, CoC should be enforced in naming conventions. 

\vspace{4pt}
\noindent \textbf{Lesson \#6: Educating developers regarding SGIDs and their implications.} A few comments in our dataset report about stereotyping or sexist behavior and the limited capability of developers to take action against those. For example, a pull request comment says, \textcolor{MidnightBlue}{\textit{``I've found that people are ignorant, ill-informed, or against diversity. So often, I encounter `there is nothing we can do,' `it is not worth the effort,' `women just don't want to work with computers,' or `women will end up having babies so ...'"}}.  This comment shows a developer being concerned about SGIDs and initiating a discussion in a pull request comment.  Education materials can be created to help developers better understand why a particular interaction is SGID and should not be used.

\subsection{Directions for future researchers:} While crafting the tool, we have gained following insights that could benefit future designers working on SGID content detection tools.

\vspace{4pt}
\noindent \textbf{ Insight \#1: Investigation of context and sexist roots of words and phrases is necessary for accurate empirical investigation. }
Understanding the sexist roots of different terms and phrases might be challenging. For example, \textcolor{MidnightBlue}{\textit{“ok soccer mom gosh”}}. Here, calling a person `soccer mom’ means that person is being called insistent and super busy. It is a stereotyping that may be acceptable in other discussions but not in the SE context. For another instance, \textcolor{MidnightBlue}{\textit{“should we really be eating power puff girl chili?”}}. Here, “power puff girl” refers to a cartoon where superhero characters are girls, and `chili' means pepper. So the person who is asking will be doing something ``difficult” based on someone’s advice (maybe a woman). 
Moreover, it is difficult to understand the intended meaning accurately without knowing the author and the target. 
For example, \textcolor{MidnightBlue}{\textit{``you are aboslut right ;)''}}. The wink emoticon at the end of the text might express that the typo is intentional and the author is trying to convey something with a different meaning.  `Karen' is another stereotyping word used for white women. We noticed a couple of instances during our labeling. Although we labeled those as `SGID's, the target person's name may be `Karen.'

\vspace{4pt}
\noindent \textbf{ Insight \#2: Domain-specific customization is necessary.} Misogyny identification is a natural language processing (NLP) task with a limited number of labeled datasets. To the best of our knowledge, ours is the first one from the SE domain. However, during our first labeling iteration, the annotators had difficulty making decisions since many phrases may convey different meanings for SE context, for example, \textcolor{MidnightBlue}{\textit{``That's good practice. \_Skinny views, fat models\_ they say."}} For a non-SE person, this text is talking about women, but in the SE domain, it discusses project components using the Model-View-Controller (MVC) architecture.  Therefore, an SE domain-specific tool is necessary. 

\vspace{4pt}
\noindent \textbf{Insight \#3: Addressing typos/ misspelled misogynistic keywords is necessary.} While filtering the contents based on keywords, we found that many texts contain sexist keywords that may have happened due to typos. For example, hoe $\rightarrow$ how, flag $\rightarrow$ fag, busy $\rightarrow$ bussy, and witch $\rightarrow$ which.
While it is difficult to discern whether those typos were intentional, we put those instances into the non-SGID group. Future researchers should keep an eye on such cases while building SGID tools. 
On the other hand, someone can alter or replace letters, for example, o with 0 or l with 1, intentionally to circumvent the automatic detection of sexist content. We used regular expression matching during the pre-processing step to identify such cases. However, these pre-processing steps also introduced several false positives. Automated identification of such adversarial examples remains a challenge.

\vspace{4pt}
\noindent \textbf{Insight \#4: More  SGIDs, especially the ones from the less represented groups, are needed to improve performance.} Despite our labor-intensive keyword list preparation, stratified sampling strategy, and labeling more than 11K instances, we found only 1,422 SGIDs. Despite these challenges, the precision and recall of our best model are close to the state-of-the-art ones achieved on Twitter or Reddit datasets~\cite{pamungkas2020misogyny}. Our error analysis suggests that lower representations from five out of the twelve SGID groups also contribute to false negatives. We noticed that 32\% of our keywords fall under the umbrella of `Pejoratives' (Table~\ref{tab:keyword_list}). Texts, including pejorative, are more likely to fall into either SO, Stereotyping, or Discredit. It might be a reason for the prominence of those three classes in our dataset. Therefore, adding more lexicons to other groups might help to find cases belonging to those. {While a multiclass classifier to identify which categories of SGID a text falls under is the ultimate goal for this research, more work needs to be done to accomplish that goal.  Since we have 12 classes, multiclass classification is extremely challenging. Moreover, our dataset is highly unbalanced, with most instances belonging to four categories. Hence, rarely represented classes would have poor performance with this dataset.  Therefore, we would require a new strategy to identify adequate instances for underrepresented classes.}

\vspace{4pt}
\noindent \textbf{Insight \#5: Customized preprocessing steps is crucial.} It is crucial to recognize that gender discrimination and other forms of toxicity and incivility come in various shapes and sizes. Therefore, customizing pre-processing steps is essential. For instance, you should employ a pattern-based matcher to pinpoint SGID-related keywords while also implementing domain-specific pre-processing steps for developers, such as splitting identifiers. Additionally, it's worth noting that incorporating optional counting-based features can potentially enhance the performance of CLE models. However, regarding DNNs and PTMs, these features might not yield the same performance boost. {We also discovered numerous texts containing sexist keywords related to a game or project scenario that we did not classify as SGID content. Future tool developers should remain vigilant for this type of material. Additionally, we encountered many false negatives in the `Discredit' and `Stereotyping' categories. More detailed preprocessing steps focused on these categories can be introduced to enhance performance.}

%% file: Sections/threats.tex
\section{Threats to Validity}
\label{sec:threats}
 \textbf{Internal validity:} 
 Our keyword-based filtering is the primary source of internal validity. We curated our list of keywords from multiple sources. We also included a keyword expansion step to identify potentially missing keywords. Yet, the lack of completeness of keywords remains a concern. We tried to mitigate this threat by randomly selecting 2,500 instances (i.e., Dataset C) that do not include our keywords. Since we did not find any SGID in Dataset C, the likelihood of missing a very large number of SGIDs due to the incompleteness of our keywords is minimal.
 Our stratified sampling using an off-the-shelf AMI tool may also introduce a bias if it performs better for certain classes of SGIDs. We mitigated this threat using a very low threshold (i.e., 0.2) and also including a sample of 2501 texts below this threshold. 
 
 \noindent \textbf{Construct validity:}
 Potential annotator biases are primary construct validity threats for this study.  Our team of annotators includes two women and two men, all aged between 20-35. However, we did not rely on the gender diversity of our team and took precautions to avoid personal biases. We followed rubrics from peer-reviewed studies~\cite{sultana2021identifying,hcissInsult,guest2021expert}.
 We annotated our rubric with examples and the annotators spent a significant amount of time discussing the rubric and talking about hypothetical positive and negative cases to build a common understanding. Moreover, we labeled the rubrics in three phases to improve their understanding based on new types of cases that they may have not seen in prior sets.
 Although a few instances from our study may still be subject to annotator biases, the number of such cases may not be large enough to invalidate our results, due to our carefully designed rubric and labeling protocol.  {
 Moreover, we mostly retained the default hyperparameters for the CLE algorithms and did not make significant adjustments. For ToxiCR~\cite{sarker2022automated}, researchers explored six parameters with a total of 5,040 combinations for RandomForest and five parameters with 360 combinations for DecisionTree and concluded that hyperparameter tuning did not lead to notable performance improvements for these algorithms. Also, due to significant computational costs, they did not conduct the hyperparameter tuning for the DNN algorithms too. Based on their findings, we opted not to perform extensive hyperparameter tuning for our SGID4SE tool.  
 }
 
 \noindent \textbf{Conclusion validity:}
 Since the primary objective of this study is to develop an automated model with a diverse set of instances, we decided to search from all GitHub communications. However, due to technical limitations such as 255 characters limit in GHTorrent and 1,000 results per query from GitHub search API,  we may have missed a large number of SGID comments, especially the ones with words appearing frequently appearing in non-SGID contexts (e.g., `girl', `mother', and `sexy'). Therefore, the distributions of SGIDs shown in Table~\ref{tab:dataset} may not be an accurate representation.  To identify more accurate distributions, we would need to carefully curate a sample of FLOSS projects on  GitHub,  download all the comments for those projects using GitHub REST API, and conduct an empirical study. With SGID4SE, such an empirical study would require less manual labeling than a keyword-based approach. Finally, duplicate instances in a dataset can provide inflated results if the same text belongs in both the training and test partitions. To avoid such pitfalls, we ensured unique instances during dataset curation. Therefore, our results do not suffer from such a threat to validity.


 

\noindent \textbf{External validity:} Selection of datasets from FLOSS projects hosted on GitHub imposes an internal threat to our study. We have curated GitHub, which does not represent the many FLOSS projects. FLOSS projects vary widely based on governance model, CoC guidelines, CoC enforcement mechanisms, and above all, community values. Therefore, this study replicated on a different FLOSS project such as Android, Chromium OS, Wikimedia, or Mozilla are likely to identify distributions among the SGID categories.

%% file: Sections/conclusion.tex
\section{Conclusion}
\label{conclusion}

Prior studies prove the manifestation of SGID content that results in push-back of women's participation in FLOSS communities~\cite{hcissInsult,singh2021motivated}. To the best of our knowledge, we conduct the first study that attempts to automatically identify SGID content from software developers' interactions. In this regard, we have built a labeled corpus of 11,007 pull request comments for SGID identification and developed SGID4SE that achieves 95.7\% accuracy with 82.8\% f1-score. We released our labeled dataset, pre-trained models, results, and summary of error analysis in the replication package. To lessen SGID content in FLOSS communities and create a more inclusive environment, administrators can adopt our dataset and tool, enhance it further, and implement it for automated identifications. 

%% file: Sections/data_availability.tex
\section*{ Data availability}\label{Data_availability}
{ We have made our labeled dataset, source code, and results publicly available on GitHub at: \url{https://github.com/WSU-SEAL/SGID4SE}}